\documentclass[iop]{emulateapj} 

\usepackage{amssymb, amsmath, amsfonts}

\usepackage{graphicx, shadow}

\usepackage{epstopdf}

\usepackage{rotating}
\usepackage{wasysym}
\usepackage{natbib}

\newcommand{\degree}{\ensuremath{^\circ}}

\newcommand{\lsun}{L_{\odot}}
\newcommand{\lbol}{L_{\rm bol}}

\newcommand{\sunrise}{\textsc{Sunrise}}
\newcommand{\gadget}{\textsc{Gadget}}
\newcommand{\arepo}{\textsc{Arepo}}
\newcommand{\gadgettwo}{\textsc{Gadget-2}}
\newcommand{\mappings}{\textsc{Mappingsiii}}

\newcommand{\tausi}{\ensuremath{r_{9.7}}}
\newcommand{\pah}{\ensuremath{r_{7.7}}}
\newcommand{\cone}{\ensuremath{f_{30}/f_{15}}}
\newcommand{\ctwo}{\ensuremath{f_{15}/f_{5.7}}}

\newcommand{\cfour}{\ensuremath{f_{4.5}/f_{1.6}}}
\newcommand{\mum}{\ensuremath{\mu m}}
\newcommand{\um}{\ensuremath{\mu m}}
\newcommand{\lagn}{\ensuremath{L_{\rm agn}/L_{\rm bol}}}
\newcommand{\lhard}{\ensuremath{L_{2-10\rm\ keV}}}

\newcommand{\mir}{mid-IR}

\newcommand{\jwst}{\textit{JWST}}
\newcommand{\jwstfull}{\textit{James Webb Space Telescope}}
\newcommand{\spitzer}{{\sl Spitzer}}

\newcommand{\thru}{\text{--}}
\shorttitle{Mid IR AGN Diagnostics}
\shortauthors{G.F. Snyder et al.}

\begin{document}
\title{Modeling Mid-Infrared Diagnostics of Obscured Quasars and Starbursts}

\author{Gregory F. Snyder\altaffilmark{1}, Christopher C. Hayward\altaffilmark{2}, Anna Sajina\altaffilmark{3},Patrik Jonsson\altaffilmark{1,4}, Thomas J. Cox\altaffilmark{5},Lars Hernquist\altaffilmark{1}, Philip F. Hopkins\altaffilmark{6,7}, Lin Yan\altaffilmark{8}}
\email{gsnyder@cfa.harvard.edu}
\altaffiltext{1}{Harvard-Smithsonian Center for Astrophysics, 60 Garden Street, Cambridge, MA 02138, USA}
\altaffiltext{2}{Heidelberger Institut f\"ur Theoretische Studien, Schloss-Wolfsbrunnenweg 35, 69118 Heidelberg, Germany}
\altaffiltext{3}{Department of Physics and Astronomy, Tufts University, 4 Colby Street, Medford, MA 02155, USA}
\altaffiltext{4}{Present Address: Space Exploration Technologies, 1 Rocket Rd, Hawthorne, CA 90250}
\altaffiltext{5}{Carnegie Observatories, 813 Santa Barbara Street, Pasadena, CA 91101, USA}
\altaffiltext{6}{Department of Astronomy, University of California at Berkeley, C-208 Hearst Field Annex, Berkeley, CA 94720, USA}
\altaffiltext{7}{Einstein Fellow}
\altaffiltext{8}{Infrared Processing and Analysis Center, California Institute of Technology, Pasadena, CA 91125}


\begin{abstract}
We analyze the link between active galactic nuclei (AGN) and mid-infrared flux using dust radiative transfer calculations of starbursts realized in hydrodynamical simulations.  Focusing on the effects of galaxy dust, we evaluate diagnostics commonly used to disentangle AGN and star formation in ultraluminous infrared galaxies (ULIRGs).  We examine these quantities as a function of time, viewing angle, dust model, AGN spectrum, and AGN strength in merger simulations representing two possible extremes of the ULIRG population: one is a typical gas-rich merger at $z \sim 0$, and the other is characteristic of extremely obscured starbursts at $z \sim 2\thru 4$.  This highly obscured burst begins star-formation-dominated with significant PAH emission, and ends with a $\sim 10^9\rm\ yr$ period of red near-IR colors.  At coalescence, when the AGN is most luminous, dust obscures the near-infrared AGN signature, reduces the relative emission from polycyclic aromatic hydrocarbons (PAHs), and enhances the $9.7\mum$ absorption by silicate grains.  Although generally consistent with previous interpretations, our results imply none of these indicators can unambiguously estimate the AGN luminosity fraction in all cases.  Motivated by the simulations, we show that a combination of the extinction feature at $9.7\mum$, the PAH strength, and a near-infrared slope can simultaneously constrain the AGN fraction and dust grain distribution for a wide range of obscuration. We find that this indicator, accessible to the \jwstfull, may estimate the AGN power as tightly as the hard X-ray flux alone, thereby providing a valuable future cross-check and constraint for large samples of distant ULIRGs.  
\end{abstract}



\maketitle

\section{Introduction} \label{s:intro}

Understanding the link between supermassive black holes (SMBHs) and their host galaxies is essential for deciphering the formation and evolution of galaxies.   Galaxy/SMBH co-evolution is expected given the observed correlations between SMBH mass and galaxy properties \citep[e.g.,][]{kr95, magorrian98, ferrarese00, tremaine02, gultekin09} and theoretical arguments that SMBH growth is self-regulated by feedback \citep[e.g.,][]{sr98,sdh05,hopkinsfp_07a,hopkinsfp_07b}.  

A key prediction from this framework is that a galaxy experiencing rapid inflow of cold gas can evolve through various classes of starbursts and active galactic nuclei (AGN), such as ultra-luminous infrared galaxies (ULIRGs) and quasars (QSOs), and that these phases are connected in an evolutionary sequence \citep[e.g.,][]{sanders88a}.  With such a model, both a starburst and a heavily obscured AGN can co-exist \citep{Hopkins:2006unified_model}.  

Testing this picture requires not only finding signatures of obscured AGN activity within starburst galaxies, but also interpreting them in the context of galaxy/SMBH co-evolution.  Locally, the most extreme starbursts are the ULIRGs \citep[$L_{\rm{IR}}\,>\,10^{12}\lsun$;][]{sanders88a}, which are almost exclusively the result of a recent/ongoing major merger \citep{sm96}, which triggers both starburst and obscured AGN activity.   To evaluate SMBH growth and the role of feedback during this critical phase, it is necessary to robustly estimate the AGN power during all observed phases.  A primary challenge to determine the fraction of flux attributable to the AGN is the reprocessing of its photons by the host galaxy's interstellar medium (ISM).  In essence, most diagnostics focus on finding signatures that are directly associated with the AGN output: narrow-line region (NLR) emission, X-rays, torus emission, or radio emission. 

A popular approach to study, in particular, dust-obscured AGN utilizes the mid-infrared (\mir) regime, loosely defined as 3-30\um, where typical starbursts have a spectral energy distribution (SED) with a minimum in continuum emission, while continuum emission from dust surrounding AGN rises toward longer wavelengths owing to hot dust emission, potentially from a torus \citep[e.g.,][]{pk92,stern05,Honig2006,nenkova08}.  

This idea led to the construction of diagnostic diagrams separating starburst from AGN dominated sources, based on Infrared Space Observatory ({\sl ISO}) data of local IR-luminous sources \citep{lutz96,lutz98,genzel98,rigo99,laurent00,tran01}.  The sensitive Infrared Spectrograph \citep[IRS;][]{houck04} aboard the \spitzer\ Space Telescope \citep{werner04} enabled the expansion of this approach to larger samples covering a wider range of physical properties \citep[e.g.,][]{jdsmith07,brandl06,schweitzer06,wu06,dale06, armus06_ngc6240,armus07,sturm06,spoon07}, and permitted \mir\ diagnostic work on objects at higher redshifts.  In particular, \mir\ diagnostics have been applied to galaxies at $z\sim1\thru3$ \citep[e.g.][]{houck05,yan07,pope08}, the epoch of peak star-formation rate density \citep[e.g.,][]{Bouwens2007} and peak quasar number density \citep[e.g.,][]{Richards2006}, which makes it a period crucial for evaluating the co-evolution of galaxies and SMBHs.  It is also when LIRGs and ULIRGs \citep[see][for a review]{sm96} make a significant contribution to the global averaged luminosity density and to the cosmic infrared background \citep{lefloch05, dole06,caputi07,Hopkins:2010IR_LF}. 

Other findings on high-$z$ IR-luminous galaxies suggest that high-redshift ULIRGs may be unlike local ULIRGs in that they may not all be late-stage mergers. For example, $z$\,$\sim$\,2 sources tend to be more starburst-dominated than $z\sim 0$ sources of comparable \mir\ luminosity \citep{fadda10}.  Differences in both the spectral energy distributions (SEDs) and morphologies of high redshift ULIRGs, as compared with local ($z\lesssim0.1$) sources, do suggest that they are not analogous \citep[e.g.,][hereafter S12]{sajina12}. Furthermore, submillimetre galaxies, a subset of high-redshift ULIRGs, may be a mix of early-stage quiescently star-forming mergers, late-stage merger-induced starbursts, and isolated disk galaxies \citep{hayward12bimod,Hayward2012}. The typically greater gas fractions \citep{Tacconi:2010} imply that fueling both star-formation and black hole accretion is relatively easier, and therefore not necessarily analogous to what we see in local ULIRGs.  

In both the local and high-redshift work, it is not clear to what extent common diagnostics miss the crucial evolutionary stage when the AGN is most deeply obscured.  It is possible that we can only detect the presence of an AGN once sufficiently many unobscured lines of sight to the AGN exist.  Moreover, while IR diagnostics generally follow the expected AGN/starburst fractions, the evolution of IR SED properties reflects a complex mix of these components which may be difficult to interpret for individual sources or samples \citep[][hereafter V09]{veilleux09}.  Therefore it is desirable to search for signposts of AGN powering that may suffer minimally from these complexities.  

To analyze this issue, numerical calculations have been used to capture the complex geometries and radiation sources relevant for disentangling AGN activity from star formation.  Originally restricted to simplified geometries \citep[e.g.,][]{Witt:1992}, calculations of dust attenuation and reprocessing are now tenable for three-dimensional dynamical simulations on galactic scales including starbursts and SMBH emission \citep[e.g.,][]{Chakrabarti:2007}.   Studies of hydrodynamical simulations with dust radiative transfer in postprocessing have found that AGN signatures, such as a power-law SED or warm IR colors in ULIRGs, are generally associated with SMBH activity \citep{Younger:2009} but may also be caused by intense starbursts \citep{Narayanan:2010dog}.  Analyzing the predicted far-IR emission from such simulations, \citet{hayward11} and \citet{hayward12bimod} found that the IR signatures of star formation vary depending on whether the activity occurs in a quiescent or bursty mode, potentially complicating discriminators of AGN and star formation activity.  One challenge raised by this body of work is how to interface the modeled SMBH with the host galaxy when it is not feasible to model the central engine fully self-consistently, a problem similar to the one described by, e.g., \citet{Jonsson:2010sunrise} for modeling young star clusters.

In this paper we combine the high diagnostic power of mid-IR features, such as PAH emission and the $9.7\mum$ silicate feature \citep[e.g.,][]{Siebenmorgen2005,levenson07}, with three-dimensional hydrodynamical galaxy merger simulations to better understand existing \spitzer\ IRS data, and to prepare for future data from the \jwstfull.  In this novel approach to AGN diagnostics, we calculate infrared SEDs from simulations using dust radiative transfer, including a simple model for AGN accretion and emission.  We then use \mir\ diagnostics to gain insight regarding AGN activity when dust in the host galaxy may be important, and also to highlight areas in which such modeling techniques might be improved.  In Section~\ref{s:simulations} we describe simulations of two representative starbursts meant to bracket the level of obscuration in ULIRGs, from which we compute and analyze the \mir\ SEDs as described in Section~\ref{s:diagnostics}.  In Section~\ref{s:results}, we show how these features depend on AGN power, evolutionary stage, viewing direction, intrinsic AGN SED, and assumptions about the dust and ISM.  In Section~\ref{s:comparison} we construct commonly used diagnostic diagrams and evaluate their ability to estimate the AGN luminosity fraction.  We consider implications of this work for future AGN modeling and studies of the evolution of IR-luminous galaxies in Section~\ref{s:implications}, and we conclude in Section~\ref{s:conclusions}.  We explore a toy model for our indicators in Appendix~\ref{appendix}.  


\section{Simulations}  \label{s:simulations}

      \begin{figure*}
	\begin{center}
	\begin{tabular}{c@{}cc}
	\includegraphics[height=3.33in,clip=true,trim = 0 -0.7in 0 0]{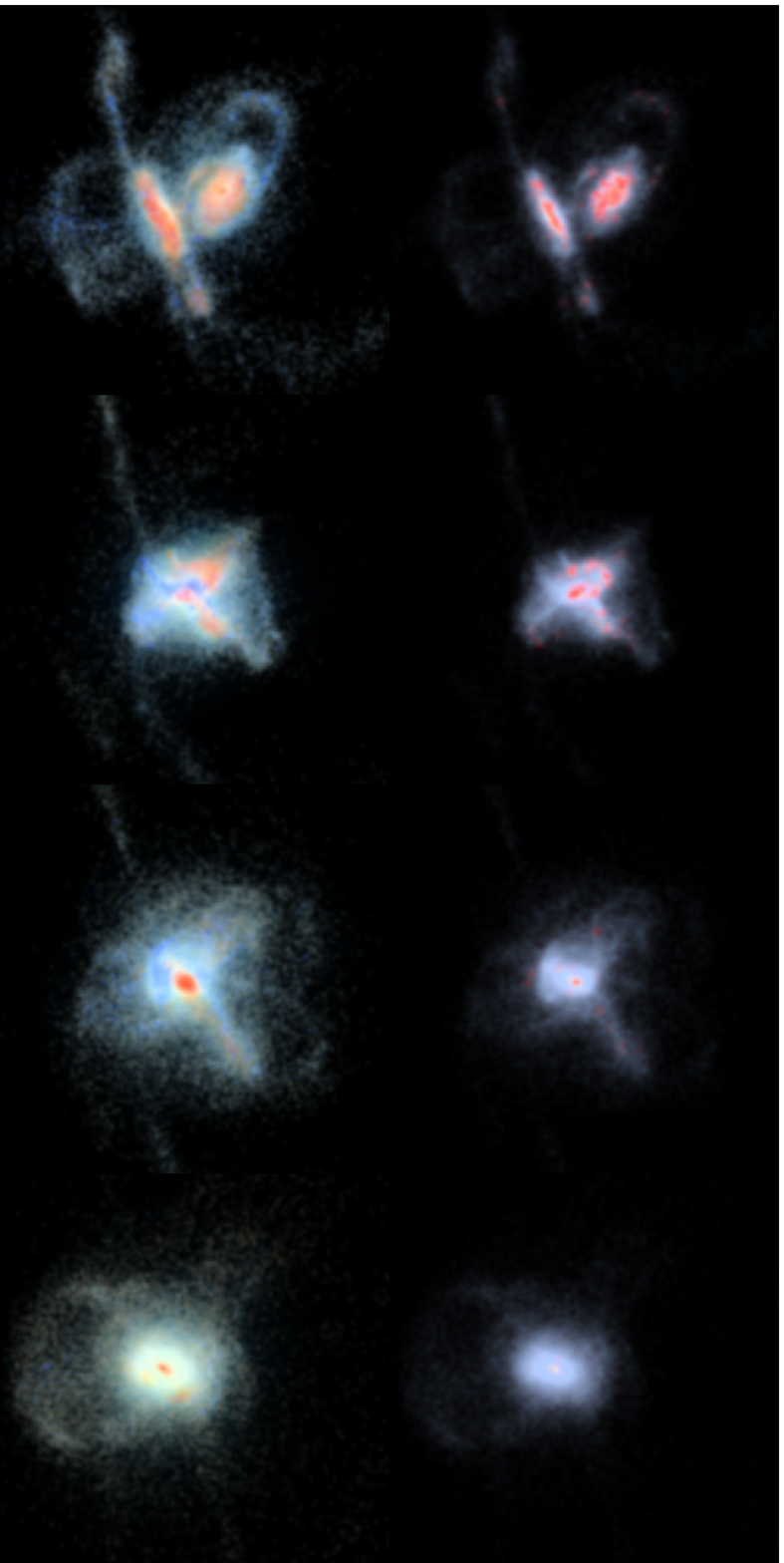} &
	\includegraphics[height=3.5in]{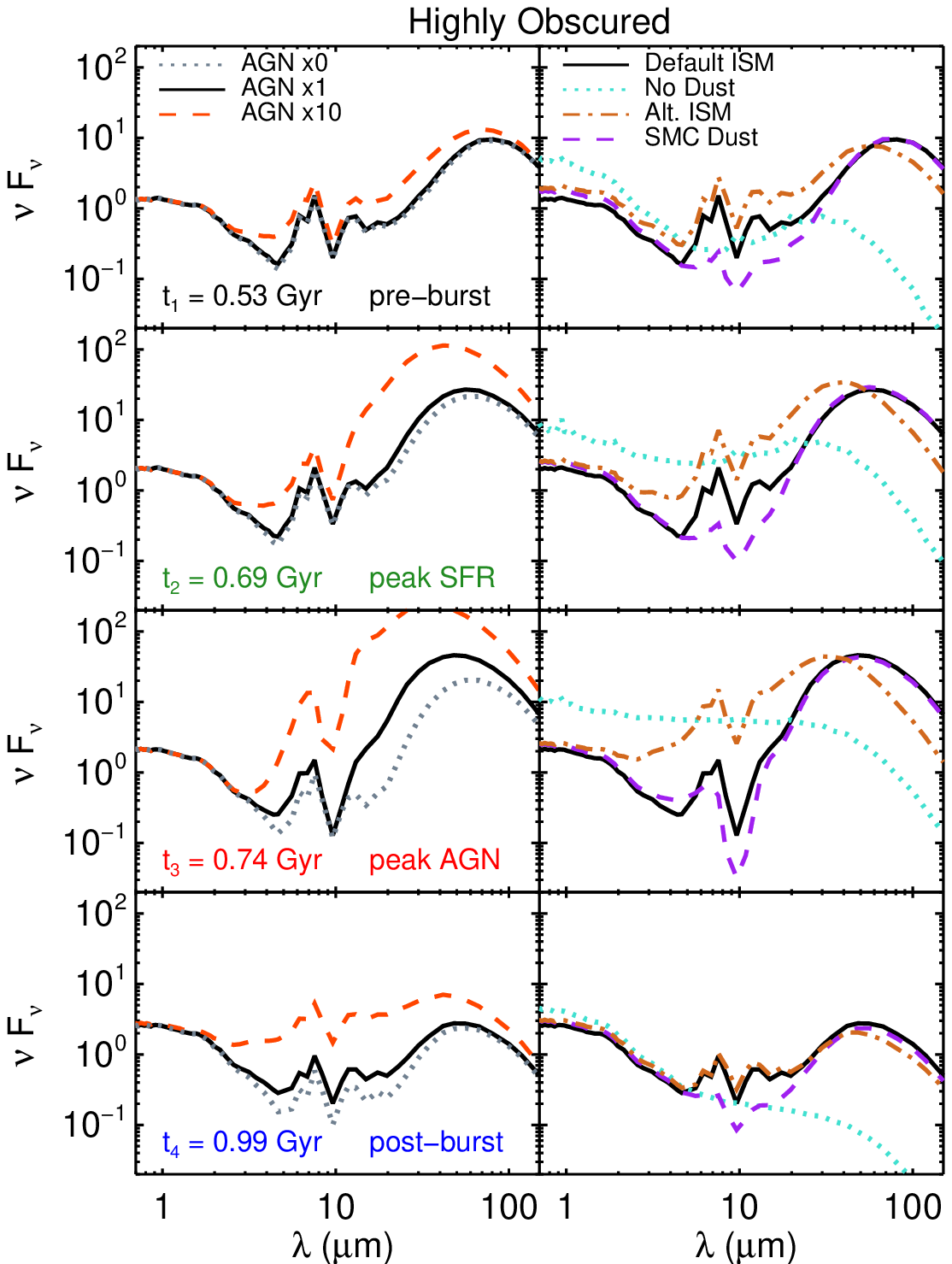} &  \includegraphics[height=3.5in]{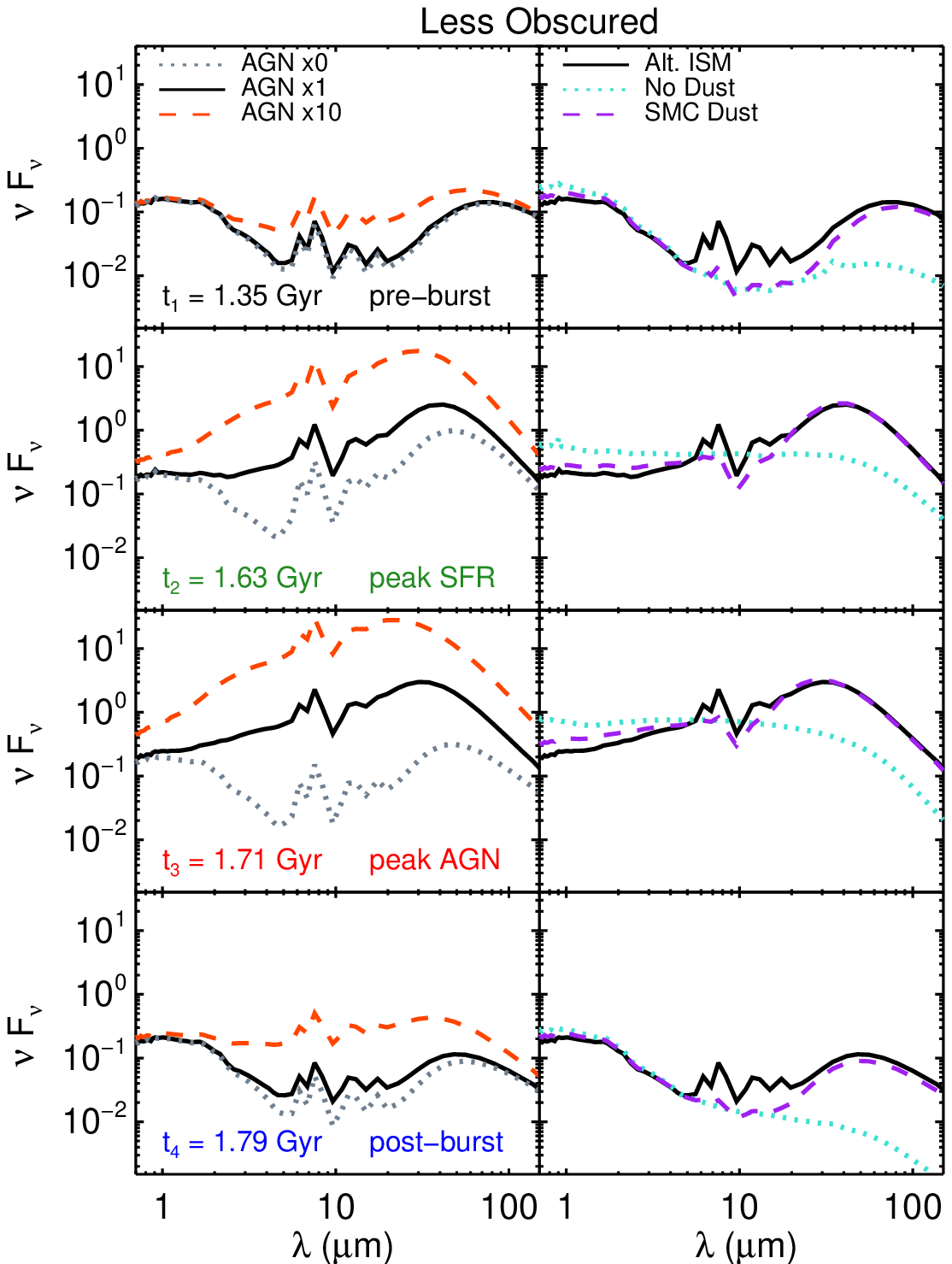}\\
	\end{tabular}
	\end{center}
      \caption{ An overview of the SEDs of our merger simulations.  On the left we show false-color rest-frame U-V-J and IRAC composite images of the highly obscured merger at each of four example times, as seen from the same direction as the SEDs to the right are measured, and utilizing our default RT model and AGN strength.  In the middle two columns we show total rest-frame SEDs of the highly obscured merger from an example viewing direction at each of four times labeled in the left-most panel.  SEDs have the same arbitrary flux normalization.  Different curves in the left column represent our fiducial dust model, but the input AGN spectrum has been multiplied by 0, 1, and 10.  Curves in the second column show our fiducial AGN level, but different ISM assumptions.  In the right two columns we similarly demonstrate the SEDs of the less obscured merger.
       \label{fig:seds}}
      \end{figure*} 

We combine high-resolution \gadgettwo\ \citep{Springel:2005gadget} 3-D N-body/smoothed-particle hydrodynamics (SPH) simulations of equal-mass galaxy mergers with the \sunrise\ \citep{Jonsson:2006sunrise,Jonsson:2010sunrise} polychromatic Monte Carlo dust radiative transfer (RT) code to examine commonly-applied mid-infrared (\mir) AGN signatures.  We focus on two representative mergers, a ``highly obscured'' hyper-LIRG analogue, and a ``less obscured'' marginal ULIRG example.  The highly obscured simulation analyzed here is the same one focused on in \citet{hayward11}.  

\subsection{Hydrodynamical simulations}  \label{ss:hydro}

\gadgettwo\ is a TreeSPH \citep{Hernquist:1989treesph} code that conserves both energy and entropy \citep{Springel:2002}.  The simulations include radiative heating and cooling as in \citet{Katz:1996}, which assumes ionization equilibrium with an ultraviolet background \citep{Faucher-Giguere2009} and no metal line cooling. Star formation (SF) is modeled to reproduce the global Kennicutt-Schmidt law \citep{Kennicutt:1998} via the volumetric relation $\rho_{\rm SFR} \propto \rho_{\rm gas}^{1.5}$ with a minimum density threshold $n \sim 0.1$ cm$^{-3}$.  The star formation law employed should be considered an empirically and physically motivated prescription to summarize physics we do not resolve.  We do not track the formation of molecular gas or resolve individual molecular clouds, and this limitation is a key uncertainty in our radiative transfer step (see Section~\ref{ss:alternatemodels}).  

The structure of the ISM is modeled via a two-phase sub-resolution model in which cold, star-forming clouds are embedded in a diffuse, hot medium \citep{Springel:2003} pressurized by supernova feedback that heats the diffuse ISM and evaporates the cold clouds \citep{Cox:2006feedback}.  All supernova energy is assumed to thermally heat the hot ambient medium, and no kinetic stellar wind kicks are used.  Metal enrichment is calculated by assuming each gas particle behaves as a closed box with a yield $y=0.02$.  Supermassive black hole (SMBH) particles accrete via Eddington-limited Bondi-Hoyle accretion and deposit 5\% of their luminosity to the nearby ISM as thermal energy \citep{sdh05,dsh05}.  The luminosity is computed from the accretion rate assuming 10\% radiative efficiency, $\lbol = 0.1 \dot{M}c^2$.

In this paper, we focus on two merger simulations, summarized in Table~\ref{tab:sims}, that represent starbursts with different characteristics.  Our ``highly obscured'' simulation is a very massive, gas-rich merger meant to mimic very luminous starbursts at $z \sim 2\thru 4$, while the ``less obscured'' simulation is more representative of typical gas-rich mergers at $z \sim 0$.  In each case, we use two identical progenitor galaxies.  The progenitors of the highly obscured merger are composed of an exponential disk with baryonic mass $4 \times 10^{11} M_{\odot}$ ($60\%$ of which is gas) and dark matter halo of mass $9 \times 10^{12} M_{\odot}$ described by a \citet{Hernquist:1990} profile. The galaxy properties are scaled to $z \sim 3$ following \citet{Robertson:2006}. The orbital parameters are identical to those of the `e' orbit of \citet{Cox:2006}.  In contrast, the progenitors of the weakly obscured example have lower mass ($M_* \sim 10^{11} M_{\odot}$), are less gas-rich ($40\%$) than the highly obscured example, and are scaled to $z\sim 0$.   Each progenitor galaxy is seeded with one SMBH particle.

\begin{deluxetable*}{ccccccccc}
\tablecaption{Properties of Merger Simulations \label{tab:sims}}
\tablehead{
\colhead{Name} & \colhead{$M_{\rm halo}/M_{\odot}$} & \colhead{$M_{\rm disk}/M_{\odot}$} & \colhead{$f_{\rm gas, init}$} & \colhead{$M_{\rm BH}/M_{\odot} (t=0)$} & \colhead{$M_{BH}/M_{\odot} (t=t_4)$} & \colhead{halo $N_{\rm part}\tablenotemark{a}$} &  \colhead{disk $N_{\rm part}$}
} 
\startdata
Highly Obscured\tablenotemark{b} & $9\times 10^{12}$     & $4 \times 10^{11}$     & $0.6$ & $1.4\times 10^5$ & $9.6\times 10^8$ & $6\times 10^4$ & $8\times 10^4$ \\
Less Obscured\tablenotemark{c}    & $2.6 \times 10^{12}$ & $1.1 \times 10^{11}$ & $0.4$ & $1.4\times 10^5$ & $8.7\times 10^7$ & $1.2\times 10^5$ & $8\times 10^4$ 
\enddata
\tablenotetext{a}{Gravitational softening lengths are $200\ h^{-1}$ pc for the dark matter particles, and $100\ h^{-1}$ pc for the disk particles}
\tablenotetext{b}{Galaxy properties scaled to $z\sim 3$ following \citet{Robertson:2006}}  
\tablenotetext{c}{Galaxy properties assumed typical for $z\sim0$}
\end{deluxetable*}

In this work we focus on signatures of AGN activity that normalize out the galaxy's mass or total luminosity, and therefore the salient difference between these two cases are the relative extents to which the merger triggers star formation, AGN activity, and obscuration.  A disadvantage with this approach is we are restricted in the range of ULIRG scenarios we can probe, owing in part to limited computational resources.   In this paper we have chosen to explore two merger simulations that potentially reflect the extremes of IR-luminous galaxies, and we discuss further consequences of this limitation in Section~\ref{ss:galaxymodels}.  

\subsection{Radiative transfer} \label{ss:rt}

We use \sunrise\footnote{see http://code.google.com/p/sunrise/ for the project source code and documentation} Version 3 in post-processing to calculate the SED observed from seven cameras distributed isotropically in solid angle every 10 Myr.  For a full description of \sunrise\ see \citet{Jonsson:2006sunrise} and \citet{Jonsson:2010sunrise}, and for a summary of components essential for this work and a description of other specific choices, see \citet{hayward11}.  We describe specific choices for some of the radiative transfer parameters in Section~\ref{ss:alternatemodels}.

\subsubsection{Sources}

{\sc Sunrise} calculates the emission from the stars and AGN in the \gadgettwo\ simulations and the attenuation and re-emission from dust.  {\sc Starburst99} \citep{Leitherer:1999} SEDs are assigned to star particles according to their ages and metallicities, and SMBH particles emit the luminosity-dependent templates of \citet{Hopkins:2007} by assuming the formula above for $\lbol$.  At \mir\ wavelengths this template emits the mean SED of \citet{richards06_sed}.  In Section~\ref{ss:torus}, we vary the AGN source SED, applying two clumpy torus models by \citet{nenkova08} that span their modeled properties.  Our templates follow typical bright AGN in that only $\lesssim 10\%$ of their flux is emitted at X-ray wavelengths

Recently formed star particles ($t < 10^6 \rm\ yr$) are assigned the \mappings\ HII region sub-grid SED models of \citet{groves08}.  These models include a dusty photodissociation region (PDR) with a tunable covering fraction ($f_{\rm PDR}$).  With a non-zero $f_{\rm PDR}$, the dust re-emission from such regions is included in the emission template.   In these models, a fraction of the carbon grains are assumed to be PAH molecules whose absorption is calculated from the cross sections by \citet{Draine:2001}.  Their emission is modeled to follow a fixed template of Lorentzian profiles designed to match \spitzer\ IRS observations of PAH emission \citep{dopita05,groves08}.  We use $f_{\rm PDR} = 0$ for the ``highly obscured'' simulation owing to concerns about the applicability of these templates to the extreme ISM densities and pressures found in this merger ; see Section 2.2.1 of \citet{hayward11} for a full discussion of these issues. 

\subsubsection{ISM Structure, Metals, and Dust}

To initialize the galaxy dust radiative transfer calculation, \sunrise\ projects the \gadgettwo\ gas-phase metal density onto a three dimensional cartesian adaptive grid.   To construct this grid, the code bounds the simulation volume with a $(200\rm\ kpc)^3$ cube and divides this initial cube into 125 equally sized subregions for the first level of refinement.  Subsequent refinement is based on an estimate for the dust optical depth in each cell from the \gadget\ metal density.  In this work, nine further refinement steps were undertaken, in which cells targeted for refinement are split into eight sectors by dividing each spatial dimension in half.  This leads to a minimum cell size of $(200\rm\ kpc/5)/2^9 = 80\ pc = 55\ h^{-1}\ pc$, roughly half of the softening length ($100\rm\ h^{-1}\ pc$) used for the hydrodynamical simulations, or $\sim 1/4$ of the SPH resolution.  For the fixed resolution of the hydrodynamical simulations, this level of refinement ensures that the SEDs are converged to within 10\% \citep{hayward11}, indicating that our radiative transfer grid settings are satisfactory for the simulations we use here.

The galaxy stars and gas were assumed to have an initial mass fraction in metals of $Z = 0.01$.  During the mergers presented here, the mean gas-phase mass fraction of metals increases to $Z \sim 0.018$ just after the merger coalescence.  Here we assume 40\% of these metals are in dust \citep{Dwek:1998}, giving a dust-to-gas ratio of $\sim 1\%$ for solar metallicity gas.  We consider the Milky Way $R_V=3.1$ (MW) and SMC bar (SMC) dust models of \citet{Weingartner:2001} updated by \citet{Draine:2007}.  These models are comprised of populations of carbonaceous and silicate dust grains inferred from observations along lines of sight to stars in the two regions.  The SMC dust grain model we apply differs from the MW model in part by having a factor of $\sim 10$ fewer small carbonaceous grains (including, e.g., the PAHs), leading to a much weaker $2175$ \AA\ absorption feature.  In active regions, small grains are expected to be depleted by a number of processes \citep[e.g.,][]{Laor1993} on short timescales, and it has been found that galactic dust appears to vary continuously between properties that are MW-like and those that are SMC-like owing to the influence of star formation activity \citep[e.g.,][and references therein]{Gordon2003}.  \citet{Draine2007} found that typical nearby spirals tend to have dust properties similar to those that give rise to the MW dust model, while bright AGN are believed to be depleted of small grains \citep{Laor1993}.  \citet{Hopkins2004} found that an SMC-like dust model better matches reddening of the optical colors in quasars.  Observations of nearby starburst galaxies were found to lack a strong $2175$ \AA\ extinction feature, also resembling SMC dust \citep{Gordon:1997,gordon98}.  Likewise, \citet{Vijh2003} found that the SEDs of $z\sim 3$ starbursts (e.g., Lyman break galaxies) indicate SMC-like dust.  

Therefore, the observed dust properties are correlated with the evolutionary stage of the galaxy, as SF and AGN activity affect the creation and destruction of grains.  We limit ourselves to specifying the dust model by hand to these MW and SMC models, but in principle an evolving model could be developed and tested against observations using a suite of simulations like those we analyze here.

\subsubsection{Radiative Transfer}

With the dust distribution and grain models set, \sunrise\ then performs Monte Carlo radiative transfer through the galaxy dust by emitting photon packets from the sources and drawing interaction optical depths from the appropriate probability distribution as the packets traverse the ISM.  For each grid cell, the temperature of each dust species (with the exception of polycyclic aromatic hydrocarbons; PAHs) is calculated assuming the dust is in thermal equilibrium, and the dust re-emits the absorbed energy as a modified blackbody.  PAH molecules are treated using a method similar to the \mappings\ models.  A fixed fraction, $f_t$, of the carbon grains with size $a < 100$ \AA\ are assumed to emit radiation thermally. The remaining fraction, $1 - f_{t}$, are assumed to be PAH molecules which emit the template spectrum from \citep{groves08}.  $f_t$ is a free parameter that we set to $f_t= 0.5$ following \citet{Jonsson:2010sunrise}, roughly matching the $8\mum$-$24\mum$ flux ratios from the \spitzer\ Infrared Nearby Galaxies Survey \citep{dale07}. In high-density regions, the dust can be opaque to its own emission, so the contribution of the dust emission to dust heating must be considered.  \sunrise\ computes this self-consistently by iteratively performing the transfer of the dust emission and the temperature calculation using a reference field technique.  

\sunrise\ calculates energy absorption by dust from radiation at wavelengths $912\rm\ \AA\ < \lambda\ < 1000\ \mum $, neglecting dust heating by radiation at energies above the Lyman limit.   This may neglect some amount of energy absorbed from X-ray radiation ($h\nu > 0.5$ keV) by atoms in dust grains.  However, since our AGN template emits $\lesssim 10\%$ of its energy at these wavelengths, and the dust may not thermalize effectively under radiation at these wavelengths, we expect that the contribution to thermal dust emission we ignore from this regime can comprise only a few percent of the bolometric luminosity \citep{Laor1993}.  Additionally, \sunrise\ neglects the effects of thermal fluctuations in stochastically heated galaxy dust grains.  

Our focus herein is the \mir\ portion of the rest-frame SEDs, $1 \text{\mum} < \lambda < 30 \mum $, a regime in which dust grains heated to a wide range of temperatures ($T \sim 100\thru1500$ K) emit thermal radiation.  Arbitrary dust temperature distributions are possible during a gas-rich galaxy merger, owing to the complex and chaotically-evolving multi-phase structure of the ISM induced during a global starburst and feedback-regulated SF and SMBH growth. Thus the accurate dust heating calculations employed by \sunrise, including treatments of multiple dust species, are essential for understanding this key phase of galaxy evolution.  

Since we will make extensive use of the 9.7 $\micron$ silicate absorption feature, we stress that the depth of the feature is determined self-consistently via the radiative transfer. The strength of the feature depends on the amount of dust and the geometry of sources (stars and AGN) and dust. For reasonable galaxy geometries we expect the AGN to be more obscured than the bulk of the stars; this differential extinction is inherently ignored when one assumes a foreground screen dust geometry, so our simulations provide an excellent way to test the uncertainties caused by this assumption.

The results of the \sunrise\ calculation are spatially resolved, multi-wavelength SEDs observed from seven directions.  The success of this approach at modeling diverse galaxy populations---both local \citep[e.g.,][]{Younger:2009,Bush:2010,snyder11a} and high-redshift \citep[e.g.,][]{Wuyts:2010,Narayanan:2010smg,Narayanan:2010dog,hayward12bimod,Hayward2012}---lends credibility to its application here.  


\subsection{Alternate Radiative Transfer Models}  \label{ss:alternatemodels}

In order to gain additional physical insight, we consider different sets of assumptions for the radiative transfer stage, including two sub-resolution models for the dust distribution on scales below that resolved by the \gadgettwo\ simulation.  For our highly obscured simulation, we will focus on the ISM treatment referred to in \citet{hayward11} as ``multi-phase off''.  This choice assumes that the dust mass contained in both phases of the \citet{Springel:2003} ISM model is distributed uniformly across each resolution element, which is likely appropriate for gas-rich mergers in which the central regions are are composed of dense, almost exclusively molecular gas.  Throughout this work we refer to this model as our ``default ISM'' treatment.  

Alternatively, we can assume that gas in the cold, dense clouds of the \citet{Springel:2003} model has negligible volume filling factor.  This choice retains the dust in the ``hot'' phase, which is assumed to have a volume filling factor equal to unity.  Specifically, dust mass that corresponds to this diffuse gas phase of the \citet{Springel:2003} model is distributed uniformly across each resolution element, while dust mass occupying the dense, cold clouds is ignored.  In this case we neglect attenuation and emission from both the cold clouds in which stars are formed and from other clouds photons encounter along the line-of-sight. This assumption thus gives a lower limit on the amount of attenuation.  We refer to this assumption as our ``alternate ISM'' treatment, and apply it to both our highly and weakly obscured simulations.  

The fraction of gas in the cold phase, and therefore the amount of dust ignored by the alternate ISM model, varies with time and position during our simulations.  For the highly obscured merger, the amount of mass in the cold phase is $\sim 38\%$ ($\sim 27\%$) at the peak of the starburst (AGN).   For the less obscured merger, this fraction is $\sim 28\%$ at the peak starburst and $\sim 15\%$ at peak AGN power.  Gas which is at higher densities contains a larger fraction in the cold phase, and so during merger coalescence gas in the central few kpc achieves a cold phase fraction of $\sim 90\%$.  Thus the alternate ISM assumption, discarding this fraction of dust in the cold clouds, more strongly affects the central regions than the galaxy as a whole.  Our radiative transfer calculation implicitly assumes that this dust does not absorb any energy, and therefore it is not included in the IR emission calculation.  This dust emission would primarily affect far-IR wavelengths, and so it should not affect the analysis presented here.  

The two treatments of subresolution dust structure we employ should be considered two plausible and physically-motivated, yet uncertain, ISM models that encapsulate unresolved processes. With current simulations it is only possible to parameterize our ignorance in this way. However, \textit{any} model used to interpret ULIRG SEDs is limited by this uncertainty.  These simulations are thus particularly useful for their ability to quantify the uncertainty caused by dust clumpiness.

For both mergers, we perform the RT calculations after multiplying the SMBH luminosity at all times by zero (AGNx0), one (AGNx1), and ten (AGNx10), allowing us to manually adjust the AGN contribution to the SEDs.  Note that the effect of AGN feedback, the thermal heating of the ISM surrounding the SMBH particles, is kept fixed at the fiducial level of Section~\ref{ss:hydro}.   The AGN contribution to the \mir\ SED arises from separate self-consistent RT calculations with each of our assigned AGN luminosities, dust grain models, and ISM assumptions.  This enables us to cleanly test how a given indicator depends on AGN luminosity for fixed geometry and galaxy ISM conditions.  At each of these AGN strengths, we apply both ISM assumptions to the highly obscured merger and the ``alternate ISM'' treatment for the less obscured case.  For these three sets of three simulations, we perform the RT using both the MW and SMC bar dust models.  

\subsection{Galaxy models} \label{ss:galaxymodels}

Although the ``high obscuration'' and ``low obscuration'' simulations are meant to mimic gas rich starbursts in massive major mergers at $z \sim 3$ and $z \sim 0$, respectively (Table~\ref{tab:sims}), these cases do not necessarily accurately reflect the real ULIRG population at a given redshift.  The ``high obscuration'' example is, by choice, a particularly massive and gas-rich merger, and thus becomes a luminous hyper-LIRG that reaches a peak $\log L_{IR} \sim 13$ briefly, having $\log L_{IR} > 12$ for $\sim 1$ Gyr.  This merger is consistent with extreme starbursts such as sub-millimeter galaxies (SMGs) and hot-dust ULIRGs at $z \sim 2\thru 4$ \citep{Hickox2012, hayward12bimod,Hayward2012}, and therefore reflects a typically observed source at these redshifts (note: this does not mean it is a typical galaxy).  By contrast, the ``weakly obscured'' example is marginally a ULIRG, briefly reaching $L_{IR} \sim 10^{12} L_{\odot}$ at merger coalescence under our default assumptions, consistent with the idea that local ULIRGs are almost exclusively ongoing mergers \citep{sm96}.  Note that the initial gas fractions assigned in our progenitors, $60\%$ and $40\%$, are high compared to comparable star-forming galaxies at these two epochs --- this is to account for our neglecting cosmological accretion and gas recycling, and a fairer comparison is to the gas fractions at the time of merger, which are $\sim30\%$ and $\sim20\%$.  

Of course, real IR-luminous galaxy samples at a given redshift will draw from a wide range of scenarios. Therefore we cannot hope to fully represent the ULIRG population in the present study.  However, we will focus primarily on the goal of separating AGN from SF activity in a small set of experiments defined by our two mergers with the radiative transfer variants described in Section~\ref{ss:alternatemodels}, whose properties may more broadly span plausible situations.  As an example, at $z \sim 3$, there are some mergers with the same stellar mass as our highly obscured simulation but a lower gas fraction.  In the Eddington-limited phase, such mergers may have higher AGN-to-SF fractions than our fiducial highly obscured calculation, so the AGNx10 model may more accurately represent such cases.  Similarly, our simulations at $z \sim 0$ may not be a close match to the conditions in all local ULIRGs.  Moreover, while the final mass of our accreting SMBHs depends primarily on the fraction of feedback energy coupled to the ISM \citep[e.g.,][]{sdh05}, the growth history of SMBH, and hence its luminosity at a given time, can depend sensitively on the adopted sub-resolution model.


Given these uncertainties, the goal of our experiments --- in particular the AGNx0, AGNx1, and AGNx10 models --- is to cover our bases and controllably boost the luminosity so that we span a wide enough range in the ratio of AGN to SF activity, and not to comprehensively predict the behavior of observed ULIRGs.  In the future, a more realistic treatment of the population may be possible as large suites of high-resolution simulations become more common.  However, many uncertainties remain not only in the physical models of the ISM and AGN accretion on sub-resolution scales, but also in how to interface emission and dust attenuation from the central engine with the host galaxy in the radiative transfer stage \citep[e.g.,][]{Jonsson:2010sunrise}.  Therefore we begin with an exploration of these few numerical experiments in a first attempt to explore the applicability of this modeling technique.  

In Figure~\ref{fig:seds} we summarize the predicted SEDs of our mergers.  We focus on four times during each merger spanning $\sim 500$ Myr, and present the total \mir\ SEDs emerging in a particular direction.  In the left SED columns we show the results of our fiducial ISM assumption, but artificially vary the total luminosity of the two (and eventually one) SMBH particles.  In the right SED columns we explore our assumptions about the ISM and dust, a study that we analyze in more detail below in Section~\ref{ss:ism} (see also Section \ref{ss:alternatemodels}).

Alongside the SEDs we show high-resolution false-color composite images corresponding to the same viewing direction as the highly obscured SEDs.  The left images present the system in the U, V, and J bands \citep{johnsonmorgan53}, and the second in the 3.6, 4.5, and 8$\mum$ channels of the \emph{Spitzer} Infrared Array Camera \citep[IRAC;][]{fazio04}.  These images provide some insight into how these systems might be classified in terms of a merger-stage diagnostic -- if this system is observed at high redshift, then the only time it is obviously a merger is at $t \lesssim t_1$.  This precedes the SMBH's peak luminosity and therefore the times at which it makes an obvious contribution to the \mir\ SED, a period spanning $\sim 1$ Gyr.

\subsection{X-ray Calculations} \label{ss:xraymethods}

We compute X-ray fluxes from SMBH particles and subsequent attenuation by the ISM following the approach described by \citet{hopkins05_qevol} and \citet{Hopkins:2006unified_model}.  This method uses the intrinsic quasar SED (Section~\ref{ss:rt} and \citealt{Hopkins:2007}), and calculates its extinction at X-ray wavelengths by applying the photoelectric absorption cross sections of \citet{morrison83}, as well as Compton scattering cross sections, each scaled by metallicity.  

A key difference between \citet{hopkins05_qevol} and the present work is their assumptions correspond to the ``alternate ISM'' treatment we described in Section~\ref{ss:alternatemodels}, where the cold clumps in the ISM, and hence often a significant fraction of the dust mass, are assumed to have a small volume filling factor.  This may be true under many conditions, but may not be applicable to extremely gas-rich ULIRGs with abundant supplies of molecular gas.  Therefore we choose to use the same assumptions for X-ray attenuation that we use for the IR calculations described in Section~\ref{ss:alternatemodels}.  For our highly obscured merger simulation we calculate the column densities that attenuate the X-ray flux both ways: by discarding the cold phase mass (``alternate ISM''), and by keeping the cold phase mass (``default ISM'').  For our less obscured merger simulation, we use only the ``alternate ISM'' model, discarding the cold phase mass.  

\subsection{AGN Fraction} \label{ss:agnfractiondefinition}

For this paper, we focus on the AGN fraction, which we define as the ratio of SMBH luminosity to total luminosity across the wavelength range used by \sunrise: $0.09 \thru 10^3 \um$.  We denote this quantity \lagn.  Our definition uses the intrinsic AGN power regardless of how it appears in the final observed SED (i.e., it is insensitive to the amount of attenuation), but it does not include AGN emission at X-ray wavelengths. This quantity, \lagn, is not normally available for a given observed sample.  Here, the SMBH accretion rate and \lagn\ are available directly from the \gadget\ and \sunrise\ calculations.  Thus we can evaluate directly the effectiveness of observed indicators, defined in Section~\ref{s:diagnostics}, at estimating the AGN contribution utilizing self-consistent calculations of the galaxy's stars, dust, and SMBH.



      \begin{figure}
	\begin{center}
	\includegraphics[width=3.2in]{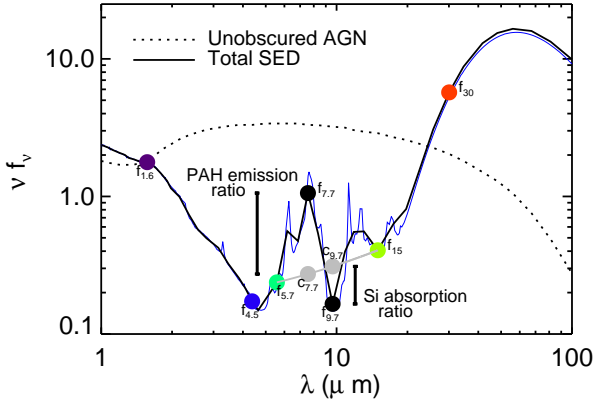}
	\end{center}
      \caption{ In the solid black line we show an example SED output from \sunrise\ in the source's rest frame.  For all subsequent analysis we use this low-resolution version, but here we plot in blue a higher-resolution calculation that demonstrates the validity of the continuum estimation that we show in gray (see Section~\ref{s:diagnostics}).  In the dashed black curve we show the intrinsic AGN spectrum used by \sunrise\ as a light source from the SMBH particles.  This SED occurs between the peak of the starburst and the peak of the AGN's bolometric contribution to the SED (between times $t_2$ and $t_3$ shown in Figure~\ref{fig:seds}).  Note that during this time the intrinsic AGN emission (attributed to the torus) at $2\um < \lambda < 10\um$ is completely absorbed and reprocessed into the far-infrared.  \label{fig:diagram}}
      \end{figure} 

\section{Mid-Infrared Spectral Diagnostics}  \label{s:diagnostics}

\subsection{From Simulations} \label{ss:simdiagnostics}

We compute mid-infrared diagnostics based on the rest-frame spatially integrated \sunrise\ SEDs, which we store as specific luminosities, i.e., the energy through the camera per unit time and per unit frequency (or wavelength) interval: $L_{\nu}$ or $L_{\lambda}$.  For simplicity and for analogy with observable quantities, we denote these as $f_{\lambda} \equiv L_{\nu} (\lambda)$, where $\lambda$ is in microns.  Owing to computational constraints, these SEDs have very low spectral resolution ($\lambda/\Delta \lambda \sim 10$), so we limit consideration to approximations of several observationally-motivated estimators.  We focus on two spectral dust features, $\tausi$, $\pah$, defined by \[\tausi =  \frac{f_{9.7}}{c_{9.7}}\] \[ \pah =  \frac{f_{7.7}}{c_{7.7}},\] where $c_{\lambda}$ is the continuum level estimated by interpolating linearly between $f_{5.7}$ and $f_{15}$.  The remaining indicators are ratios of $f_{\lambda}$ at $1.6$, $5.7$, $15$, and $30\ \mum$.  These points were chosen to match common mid-IR colors, in which a point is often chosen just below $6\um$ to avoid the $6.2 \mum$ PAH and $6\mum$ water ice features.  Unless otherwise specified, we present all SEDs and their derived quantities in the source's rest frame.  

In Figure~\ref{fig:diagram} we present an example rest-frame SED in the \mir\ and highlight the spectral features that we will consider here.  In addition to the final SED, we plot the intrinsic SMBH SED that we assume, which exhibits a generally flat shape with an ``IR bump''.  If this source were observed unobscured, or obscured only by an optically thin (at several $\mum$) dust column, then this would resemble a ``power-law-like'' AGN source with red near- and mid-IR colors.  We see this situation in the $t_2$ and $t_3$ panels for the less obscured merger (right side) of Figure~\ref{fig:seds}:  the $\cfour$ slopes depend strongly on the AGN strength, and a comparison of the ``no dust'' and ``alternate ISM'' cases indicates that dust attenuation is insignificant at wavelengths longer than a few microns.  Furthermore, a near-IR slope (such as $\cfour$) has been used to select AGN in wide-field IR surveys \citep[see, e.g.,][]{stern05,stern12}.  Thus we will keep this feature in mind while analyzing the \mir\ properties of more highly obscured sources.

By contrast, the starburst SED falls as lambda increases at $1\mum < \lambda < 10\mum$, but also exhibits distinctive features from dust grains identified as polycyclic aromatic hydrocarbons (PAHs) superimposed at 3.3, 6.2, 7.7, 8.6, 11.3, 12.6, and 17\um.   For example, SEDs characterstic of starbursts are seen in the first row of SEDs in Figure~\ref{fig:seds} (i.e., the black curves at time $t_1$) .  

\subsection{From Observations} \label{ss:observedspectra}

Observationally, weaker PAH and stronger continuum (i.e., lower PAH equivalent width (EW)) effectively indicate a higher relative contribution from the AGN in dusty galaxies, especially ULIRGs \citep[e.g.,][]{laurent00,sturm00,tran01}. This diagnostic is supported by mid-IR fine structure line diagnostics implying an AGN-like radiation field in sources of lower PAH EW \citep{genzel98,armus07}.  

More recently, such \mir\ diagnostic studies discovered sources that have a small PAH EW but a deep 9.7\um\ silicate absorption feature \citep[e.g.,][]{houck05}, which seems to require a deeply embedded, centrally concentrated source \citep{levenson06}.  These sources are believed to represent Compton-thick AGN \citep{bauer10,georg11}, although there is at least one known star forming dwarf without an AGN whose mid-IR spectrum fits the above criteria \citep{roussel03}.  

In Section~\ref{s:comparison}, we will compare our predictions with data for local and high-$z$ starbursts with available low-resolution \spitzer\ IRS spectra. Specifically, we use the ULIRG sample of V09 and a sample of 192 24\um-selected $z$\,$\sim$\,0.3\,--\,3 starbursts and obscured quasars. The redshifts and IRS spectra of the 24\um-bright sample are presented in \citet{yan07} and \citet{Dasyra2009}, while the full IR SEDs are compiled in \citet{sajina12} (S12).  

Using the IRS spectra for both ULIRG samples, we estimate $r_{7.7}$, $r_{9.7}$, $f_6$, $f_{15}$, and $f_{30}$ in the same manner as for the simulated SEDs.  However, the observed spectra have a more limited spectral coverage.  The observed coverage of the IRS spectra varies between 5.2 and 38\um, or 14\,--\,38\um\ if only the LL modules are used, as is the case for most of the higher-$z$ sample.  This means that for the local ULIRG sample, the required rest-frame 6-30\um\ is covered by the available IRS spectra, while for the 24\um-bright sample, the 15\um\ and 30\um\ continuum points are outside the IRS coverage above $z$\,$\sim$\,1.5. Below $z$\,$\sim$\,1.5, the 6\um\ continuum point is often outside the IRS coverage (depending on whether or not the SL module is available), and even the 7.7\um\ point can be outside the IRS coverage below $z$\,$\sim$\,0.8.  We compensate using the empirical SED model fits from \citet{sajina12}, which necessarily introduces additional systematic uncertainty.  These extrapolations are constrained below or above the IRS spectral coverage with IRAC 3.6, 4.5, 5.8, and 8\um, and MIPS 70\um\ photometric points, ensuring that the mid-IR colors obtained are reasonably accurate. 

We choose to exclude the $z$\,$\lesssim$\,0.9 sources from this sample in order to avoid all cases where the 7.7\um\ PAH feature is not covered by the IRS spectra. This step also removes all sources whose overall IR luminosities ($L_{3-1000}$) place them in the LIRG rather than ULIRG category.  We also exclude the $z$\,$>$\,2.5 sources, for which the 9.7\um\ silicate absorption feature is too poorly defined, being only partially covered by the IRS spectra.  This leaves a total of 118 sources.  We explore determining the above quantities with and without interpolating the IRS spectra onto the much lower resolution simulated SED wavelength array,c and we find no significant difference in the overall results.  

Lastly, in 4 cases, the silicate feature is saturated making the measured $r_{9.7}$ value a lower limit. The saturation flag is triggered when the mean of the IRS spectra between 9.0 and 10.4\um\ is less than or equal to the standard deviation in the same region.  

In Section~\ref{s:implications} we discuss the rest-frame near-IR as another means of diagnosing AGN power.  For observational comparisons in this regime, we use \spitzer\ IRAC photometry for the high-redshift sample, and for the low-redshift ULIRGs, we interpolate from 2MASS H and K band photometry (using total magnitudes).  The unknown aperture corrections add some uncertainty here; however, ULIRGs are typically compact enough in the near-IR relative to the $\sim 2.5\rm\ arcsecond$ 2MASS beam that aperture effects should be small \citep{Surace1999,Surace2000}.  For several sources, we use $2.5\thru 5\mum$ spectra \citep{Imanishi:2008,Sajina2009} from AKARI for more accurate near-IR fluxes.


      \begin{figure}
	\begin{center}
	\includegraphics[width=3.4in]{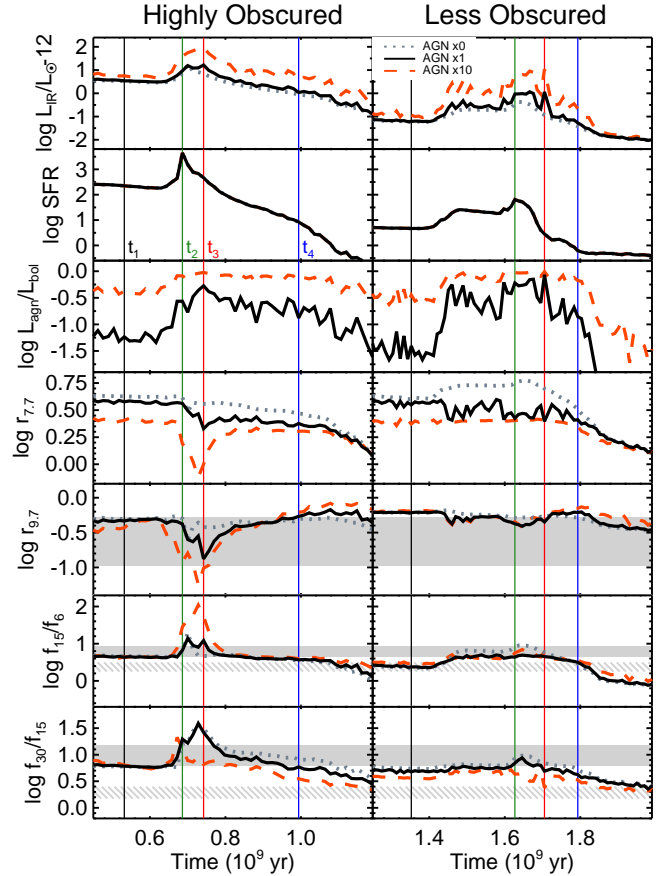}
	\end{center}
      \caption{We compare various rest-frame \mir\ quantities as a function of time for our highly obscured (left) and less obscured (right) starburst simulations.  At each time, we plot the median over viewing angle of each quantity.  Vertical lines highlight the four characteristic times $t_1\thru t_4$ from Figure~\ref{fig:seds}, corresponding to pre-burst, peak starburst, peak AGN, and post-burst.  The three curves in each panel represent the same models shown in the left column of Figure~\ref{fig:seds}: we multiply the AGN luminosity deduced from the simulation by 0, 1, or 10.  In several rows, the gray solid (lined) bands subtend the 25-75$\%$ range in the distribution of local ULIRGs (quasars) compiled by \citet{veilleux09}.  In the highly obscured case, the AGN is most dominant and most obscured at the time $t_3$, when the PAH emission ($\pah$) and Si absorption ($\tausi$) ratios are minimized, in agreement with previous reasoning.  The mid-infrared colors $\cone$ and $\ctwo$ also exhibit a weaker signal around this time, but this signal can be confused with the starburst phase ($t_2$) for $\ctwo$, and depends non-linearly or not at all on the AGN strength with $\cone$.  The more typical starburst represented in the right column is a marginal ULIRG, and its mid-infrared SED is relatively insensitive to the presence of the AGN which dominates at $1.4 < t/{\rm Gyr} < 1.8$. \label{fig:time_evolution}}
      \end{figure} 

\section{Simulation Results}  \label{s:results}


\subsection{Time Evolution}   \label{ss:agn}

In Figure~\ref{fig:time_evolution}, we explore how the mid-IR diagnostics vary with time for our two fiducial simulations: ``highly obscured'' and ``less obscured''.  We show spectral diagnostics along with quantities of interest such as the overall IR luminosity, the star formation rate (SFR), and the AGN fraction, \lagn.  We assume our default ISM model and default AGN torus model, but we show the effect of varying these choices in Section~\ref{ss:ism} and Section\,\ref{ss:torus}, respectively.  All curves shown are the median over the viewing angle. The scatter introduced by the uncertainty in the viewing angle is addressed in Section~\ref{ss:angle}.

To guide the discussion, we indicate the four times from the panels in Figure~\ref{fig:seds} with vertical lines for the highly obscured (left column) and less obscured (right column) mergers. The progenitors are still separated at time $t_1$, which precedes the period of final coalescence and peak $L_{IR}$ by $\sim 200$ Myr.  The maximum SFR and AGN power occur at $t \sim t_2\thru t_3$, after the galaxies have merged, but the central sources are still largely obscured by dust.  At $t_4$, up to $\sim 300$ Myr later, the SFR has plummeted but the AGN fraction is still elevated, and the remnants are on their way to being ellipticals.

Qualitatively, we find that our simulated mid-IR diagnostics behave largely as expected, but not always. Below we address each of the diagnostics in turn.

The PAH strength, $r_{7.7}$, drops at the time of coalescence ($t_2\thru t_3$), reaching its lowest values at $t_3$ when the AGN fraction is highest.  Pre-coalescence, the PAH strength is weakest for the model with strongest AGN (AGN$\times$10). Post-coalescence, the PAH strength again increases but does not return to its pre-coalescence levels. No significant ``coalescence'' dip in the PAH strength is seen in the less obscured case.  The level of absorption by silicate grains, $\tausi$, is greatest when the AGN is most luminous, which is also when the relative strength of the PAH emission is weakest.  These extrema are relatively sharp and confined to the period of final coalescence at $\sim t_2$-$t_3$, although the AGN luminosity fraction remains elevated for $\approx 1\rm\ Gyr$.  Overall, the shape of the PAH strength curve in both the highly obscured and less obscured cases does not bear a strong resemblance to the AGN fraction curve nor to the SFR curve.

In the highly obscured case, the $\ctwo$ slope is reddest when the AGN is most luminous, a somewhat counter-intuitive result since the 6\um\ continuum is generally believed to be enhanced by AGN torus emission. This result likely owes to the fact that the time when the AGN fraction is greatest corresponds to the time of greatest obscuration, when the 9.7\um\ silicate absorption feature is deepest. This obscuration would lead to significant absorption of the observed 6\um\ continuum, reddening the $\ctwo$ color.  However, this may result from the simulations not having sufficiently clumpy ISM to allow for lines of sight directly to the AGN torus.  This slope is also reddened at $t_2$, the time of peak SFR activity, as expected due to enhanced HII-region type emission.  For the less obscured case, the $\ctwo$ color is reddened throughout the merger ($t_1\thru t_4$), likely the result of the enhanced SFR, but lacks a sharper reddening at the time of coalescence.  This supports our view that the ``coalescence'' reddening seen in the ``highly obscured'' case is a direct result of galaxy dust attenuation.

Similarly, the $\cone$ slope is effectively constant for the duration of the merger for the less obscured case, but is significantly reddened at the time of coalescence in the highly obscured case.  At a given time, this slope is bluer when the AGN source is more luminous (e.g., AGN$\times$10), and in contrast to $\ctwo$, $\cone$ does not experience a sharp peak at $t_{3}$ for the most powerful AGN in the highly obscured merger.

Overall, our simulations indicate that the dependence of any of these mid-IR spectral diagnostics on the AGN fraction is time-dependent and non-linear and is affected by the properties of the host galaxy.  The dependence on the overall level of obscuration, as parametrized by the silicate absorption depth, is to some degree stronger.  For example, these indicators vary much less when there is less obscuration.

When the highly obscured and less obscured simulations have the same AGN fraction, the highly obscured case shows much sharper coalescence-stage SED reddening and deepening of the silicate feature. These effects on the SED are therefore the direct result of the host galaxy dust attenuation. This supports earlier observational evidence \citep{Lacy2007,Sajina2007,Juneau2011} that in at least some AGN-dominated, deep silicate absorption sources, the absorption could arise in the host galaxy rather than the AGN torus.

     \begin{figure}
        \begin{center}
        \includegraphics[width=3.4in]{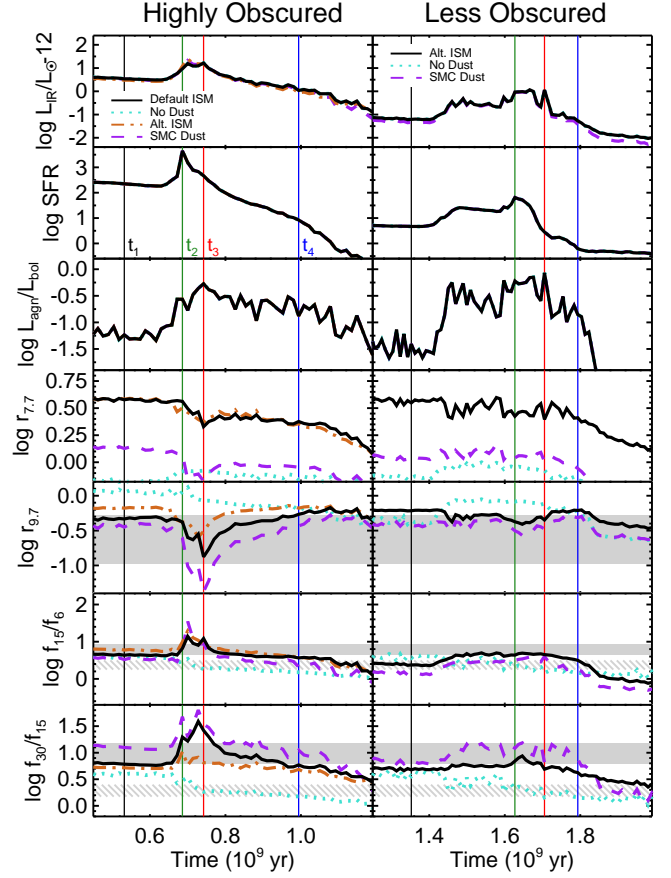}
        \end{center}
     \caption{ We compare the dust models in the same simulations and quantities as Figure~\ref{fig:time_evolution}.   For both bursts, we show our fiducial models in solid black curves, the SMC dust grain model in the dashed purple curves, and the fiducial models without galaxy dust in the light blue curves.  Vertical lines highlight the four characteristic times $t_1\thru t_4$ from Figure~\ref{fig:seds}, corresponding to pre-burst, peak starburst, peak AGN, and post-burst.  In the deeply obscured case, we also plot two alternate models: the SMC dust grain model and the clumpy ISM structure model.  The powerful AGN signal in $\tausi$ and $\pah$ at $t_3$ (Figure~\ref{fig:time_evolution}) can be replicated, at normal AGN strength, with the silicate-heavy SMC dust grain distribution.  In several rows, the gray solid (lined) bands subtend the 25-75$\%$ range in the distribution of local ULIRGs (quasars) compiled by \citet{veilleux09}. \label{fig:time_evolution_dust}}
     \end{figure}


\subsection{Dust Model Dependence} \label{ss:ism} \label{ss:dust}

In this section, we address the effect of varying both the dust structure in the ISM (clumpy vs. smooth) and its composition (Milky Way-like vs. SMC-like).  In Section 2.4, we described the two ISM treatments we apply to the highly obscured merger.  Broadly speaking, the ``default ISM'' case assumes the dust and gas associated with both hot and cold ISM phases are smoothly distributed.  The ``alternate ISM'' assumes the cold phase gas is in clumps sufficiently dense to have a negligible volume filling factor, so that the dust it contains does not contribute to the overall attenuation.  In Figure~\ref{fig:time_evolution_dust}, we explore these effects on the \mir\ SEDs.

When switching to the alternate ISM assumption for the highly obscured simulation, the magnitude of $\tausi$ is reduced by $\sim 0.3$ dex, and the color $\cone$ by $\sim 0.7$ dex, while $\pah$ and $\ctwo$ are mostly unchanged.  In addition, the total IR luminosity (first row of Figure~\ref{fig:time_evolution_dust}) is negligibly affected by this change.  Therefore the differences from this ISM structure assumption arise only because the source energy is redistributed in a different way, and not to changes in the total amount of energy that is attenuated.  This implies that there is less dust self-absorption in the alternate ISM model, which typically experiences less attenuation than the default case, so the source radiation that is reprocessed into the near-IR and mid-IR is not as often reprocessed again to longer wavelengths.  

With SMC or MW dust, the IR SEDs at a given time during the starburst have similar slopes $\cone$ and $\ctwo$ (see also, Figure~\ref{fig:seds}).  Furthermore, $L_{IR}$, a measure of the amount of attenuated light, is nearly identical in these cases.  By contrast, the SEDs show very different spectral characteristics $\pah$ and $\tausi$, reflecting the difference between MW and SMC grain composition.  The SMC dust is assumed to contain relatively more silicate than carbonaceous grains, leading to virtually non-existent PAH emission $\pah$ and correspondingly stronger silicate absorption $\tausi$.


     \begin{figure}
        \begin{center}
        \includegraphics[width=3.2in]{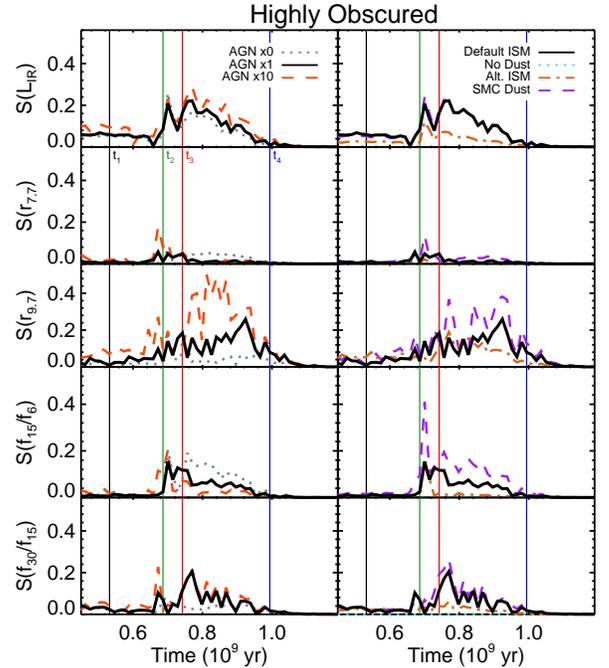}
        \end{center}
     \caption{ Scatter S, in dex, between viewing angles for the infrared quantities of the highly obscured case from Figures~\ref{fig:time_evolution} and \ref{fig:time_evolution_dust}.  S is defined in the text of Section~\ref{ss:angle}.  Vertical lines highlight the four characteristic times $t_1\thru t_4$ from Figure~\ref{fig:seds}, corresponding to pre-burst, peak starburst, peak AGN, and post-burst.  The scatter S is largest when the AGN is strongest, $\sim t_3\thru t_4$, because the obscuration varies significantly among the different sightlines to the AGN.  The camera angle scatter for the less obscured burst (not shown) is much smaller, with S always below 0.1 dex, and its median value below 0.05 dex.
\label{fig:angle}}
     \end{figure}

\vspace{0.3in}  

\subsection{Viewing Perspective Dependence} \label{ss:angle}

In addition to the (evolving) source SEDs, the observed SED depends on the evolving relative distributions of sources and dust.  This evolution changes the relative obscuration of AGN and stellar sources, affecting the behaviors of mid-IR indicators depending on whether the luminosity is dominated by star formation or AGN, because the AGN is generally much more highly obscured than the stellar component.  In Appendix~\ref{appendix}, we use toy models to confirm and further examine the behavior of the simulated diagnostics.  

Figure~\ref{fig:angle} shows the time-dependence of variation in the \mir\ quantities with viewing angle, focusing on the highly obscured merger.  The camera angle scatter is much lower (median $S < 0.05$ dex) for the less obscured merger.  We show the same assumptions regarding AGN strength and ISM models from Figure~\ref{fig:time_evolution_dust}, and we plot the scatter S, in dex, of the \mir\ quantities as a function of time.  We define $\text{S(q)}$ to be the normalized median absolute deviation \citep[MAD, e.g.,][]{mostellertukey77, beers90} of the quantity q among the seven \sunrise\ viewing directions at a particular time during the simulation.  First we calculate the median $M_q$ of q, and then the seven values $|q-M_q|$, the median of which we define to be the MAD of q.  We define $\text{S(q)} = \text{MAD}/0.67$ so that if q is drawn from a Normal distribution with standard deviation $\sigma$, then S is an unbiased estimator of $\sigma$.

At $t \lesssim t_2$, $S < 0.1$ dex for all models, so the SED shape and spectral features are similar in all directions.  Generally speaking, the scatter S reaches a maximum ($\sim0.2\thru0.4$ dex) for conventional AGN diagnostics during $t_3$ to $t_4$.  These times are also when the AGN luminosity fraction and IR luminosity are highest.  Thus the increased viewing-angle scatter implies a large AGN contribution to the IR SED.  This is because the AGN is effectively a point source, so the observed SED is sensitive to attenuation along a single LOS, which can vary widely.  Moreover, the sudden change in the extinction curve at $9.7\mum$ causes $\tausi$ to have an exagerrated dependence on the line of sight attenuation.  $\pah$ experiences much less viewing angle scatter than $\tausi$, $S(\pah) \sim 0.0$ versus $S(\tausi) \sim 0.2$ dex, because the extinction is the same for both the PAH features and $7.7\mum$ continuum.  Turning off AGN emission leads to essentially no viewing angle variation in either quantity.  

\begin{figure}
        \begin{center}
        \includegraphics[width=3.2in]{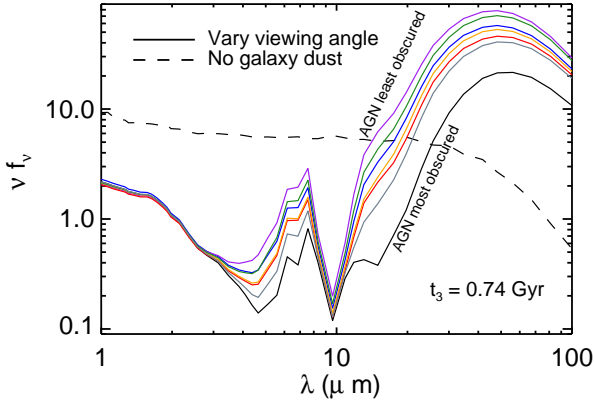}
        \end{center}
     \caption{  Infrared SEDs of the $\rm AGN\times1$ highly obscured burst simulation
at time $t_3$ (AGN peak) observed from each of the seven viewing angles.  The
IR spectrum varies owing to the anisotropy of extremely optically
thick material surrounding the central engine (starburst or AGN).
Along a given direction, this dust absorbs IR light and reemits it
into other lines of sight.    \label{fig:cameraseds}}
     \end{figure}

Figure~\ref{fig:cameraseds} presents the SED from seven viewing angles at time $t_3$ (peak AGN) for our highly obscured merger, with $\rm AGN\times 1$ and fiducial ISM models.  This gives us a clearer physical picture of what is driving changes to the \mir\ diagnostics during coalescence.  The generally higher level of obscuration at these times leads to the enhanced silicate absorption, causing the dip in median $\tausi$ at $t_3$ (Figure~\ref{fig:time_evolution}).  However, at a fixed time during this period of very high attenuation, like the one shown in Figure~\ref{fig:cameraseds}, the silicate absorption feature is stronger along lines of sight in which the AGN is less obscured.  In other words, the silicate feature only correlates positively with AGN attenuation until the level of attenuation is so extreme that the emergent stellar SED dominates the mid-IR emission associated with the AGN.  This phenomenon is derived in Appendix~\ref{appendix} for a toy model of $\tausi$.

\subsection{Intrinsic AGN Emission Dependence} \label{ss:torus}

\begin{figure}
	\begin{center}
	\includegraphics[width=3.2in]{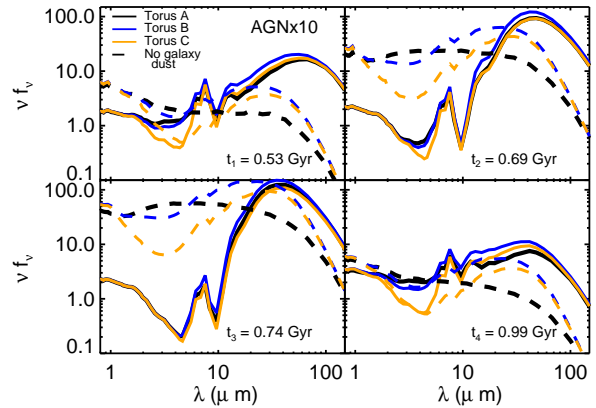}
	\end{center}
      \caption{ We explore the sensitivity of the final infrared SED to the assumed intrinsic AGN (or AGN+torus) spectrum, focusing on the highly obscured simulation.  In this figure we use the manually boosted AGN (x10) input SED, and so the effect sizes shown here should be considered upper limits.  Solid lines show the emergent SED from an example viewing direction with three assumptions about the intrinsic AGN SED.  Dashed lines show the intrinsic stellar+AGN source emission before the dust RT calculation.  Black lines use the mean templates from \citet{richards06_sed}, while blue and yellow lines use $i=0\degree$ and $i=90\degree$ clumpy torus models from \citet{nenkova08}, respectively.  The choice of intrinsic AGN SED matters least at $t_3$, when the SMBH's contribution to the total luminosity is largest.  The AGN fraction at times $t_2\thru t_4$ are all $\gtrsim 70\%$ and yet the appearance of the \mir\ SED changes dramatically over these several $10^8\rm\ yr$.  The influence of the intrinsic AGN spectrum on the final SED at $5\mum$ increases from $\sim 0$ dex at $t_3$ to $\sim 1$ dex at $t_4$.  \label{fig:torus}
      }
      \end{figure}

In Figure~\ref{fig:torus} we consider the effect of intrinsic \mir\ AGN SED shape on the final observed SED.  We show the manually boosted model (AGN$\times$10) in order to maximize the possible effect.  We demonstrate that the intrinsic AGN mid-IR spectrum has no impact on the observed mid-IR SED at $t_2\thru t_3$ of the highly obscured merger, during the starburst and AGN peak at final coalescence when the central densities are highest.  We use our default AGN SED, the template by \citet{Hopkins:2007}, as well as two templates by \citet{nenkova08}.  These templates reflect obscuration by clumpy tori inclined by $i=0\degree$ and $i=90\degree$, spanning the range of derived \mir\ torus SED shapes.  

With our default ISM assumption, dust obscures this central source at wavelengths as long as $\lambda \sim 20\thru 50\mum$.  Thus the flux at $\lambda \sim 2-5\mum$, conventionally attributed to AGN tori, only indicates direct AGN emission when the central source is sufficiently unobscured (e.g., $t_1$ and $t_4$).  In the less obscured simulation, the AGN torus SED has a large effect throughout the merger, as seen by the near-IR slopes of the SEDs in the left column of the ``Less Obscured'' panels in Figure~\ref{fig:seds}.  


      \begin{figure*}
	\begin{center}
	\includegraphics[width=6.0in]{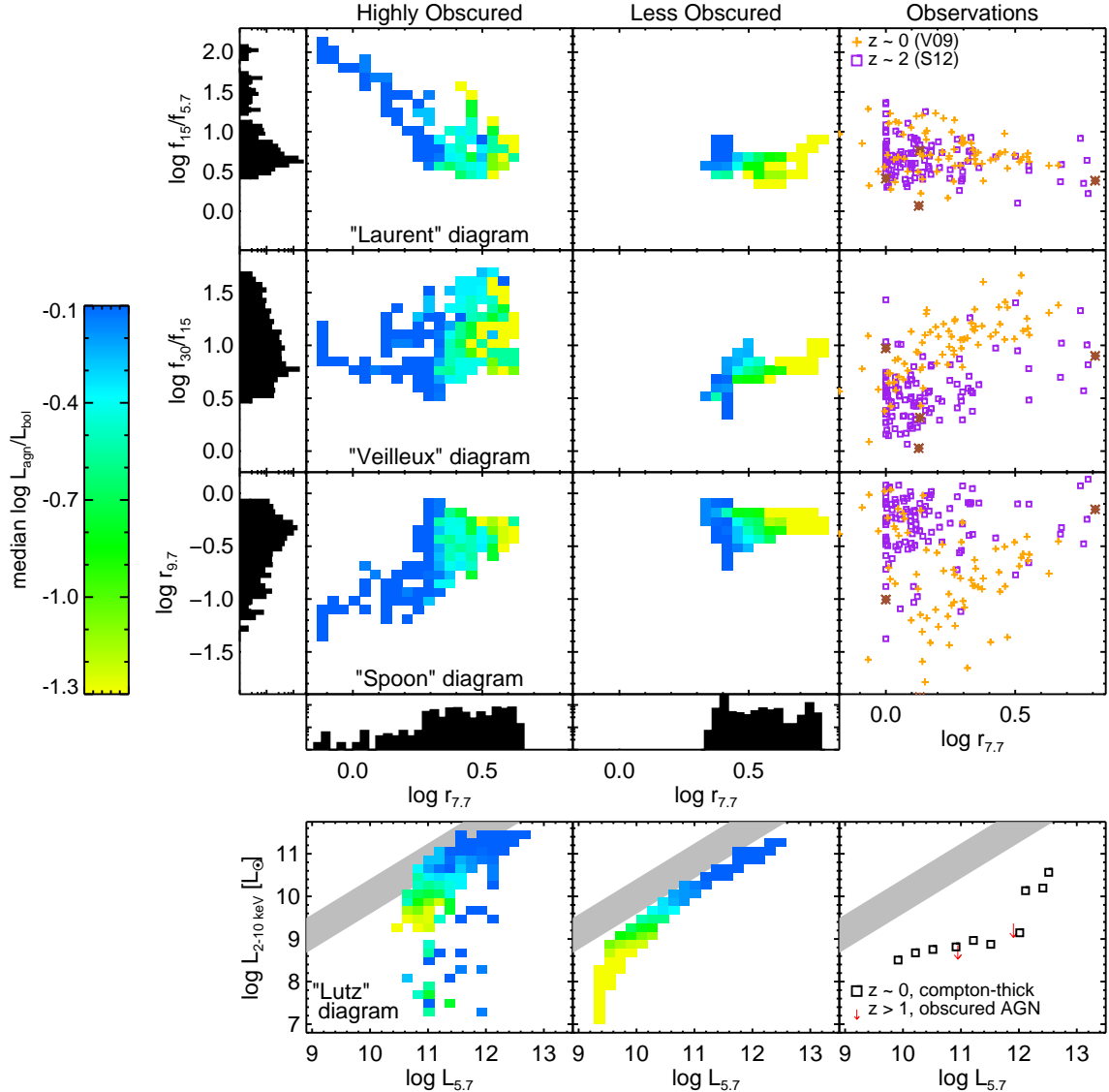}
	\end{center}
      \caption{ We compare the two simulations to observed samples of ULIRGs.  The four rows each present an example diagnostic diagram that has been examined in the literature.  Within each row, the left plot shows the ``highly obscured'' model points colored according to the median AGN fraction \lagn\ within each grid cell: see the text for a description of this procedure, and refer to the colorbar at the left.  The middle panel of each row shows the same mapping for the ``less obscured'' simulation.  We employ the fiducial ISM settings for these panels, in particular MW dust, and include all points between $t_1\thru t_4$ from the AGNx10, AGNx1, and AGNx0 models as a means of exploring the sensitivities of these SEDs to changes in modeled AGN strength.  We show a histogram around the edges of the first three columns to highlight the relative amount of time spent by the simulations in each region.  We find that dark squares span almost the entire y axis in the left column of the top three rows, implying that powerful AGN can occupy any value of these diagnostics in our two merger models. The right panel in each row plots observed local V09 and high-z S12 ULIRGs.  When the $\tausi$ feature is saturated, we plot a brown asterisk.  In the ``Lutz diagram'' in the bottom row, the gray shaded region shows the $1\sigma$ range of intrinsic X-ray-to-mid-IR ratio for Type 1 AGN from \citet{bauer10}.   \label{fig:data_comparison}}
      \end{figure*} 

\section{Observational Comparison} \label{s:comparison}

\subsection{2-Dimensional Diagnostics} \label{ss:diagrams}

In Section~\ref{s:results}, we presented the time evolution of the observed mid-IR indicators and noted that their dependence on \lagn\ is both time-dependent and non-linear.  We showed that our \mir\ spectral diagnostics strongly depend on the dust model and can be highly scattered as a function of veiwing angle.  How could we hope to observationally discern the level of AGN in a given galaxy based on its mid-IR spectrum, given that typically no information on the merger stage is available, the dust model is even less constrained, and we will never have more than one viewing perspective?  Here we attempt to do just that by using the simulation results as a diagnostic tool.  

In Figure~\ref{fig:data_comparison}, we present \mir\ diagnostic diagrams that have been used in the literature for the purpose of disentangling the role of AGN and starburst activity in dusty galaxies.  Specifically, using our definitions of the PAH strength, silicate feature depth, and \mir\ colors, we present slightly modified versions of the ``Laurent'' diagram \citep{laurent00}, the ``Spoon'' diagram \citep{Spoon2004}, the ``Veilleux'' diagram \citep{veilleux09}, and the ``Lutz'' diagram \citep{lutz04}.  The last diagram presents the relation between the $6\mum$ luminosity and X-ray luminosity for unobscured or mildly obscured AGN. The left-hand column of Figure~\ref{fig:data_comparison} presents the highly obscured simulation, the middle column presents the less obscured simulation, and the right-hand panel presents real galaxy data that we discuss in more detail in the following section.

For the simulation columns in Figure~\ref{fig:data_comparison}, we plot the IR quantities in the following way.  Each panel is divided into a grid of 25x25 bins, each of which contains zero or more simulated points.  These points include all seven viewing angles at each timestep in $t_1\thru t_4$, which are sampled every $10^7$ years.   We use our fiducial ISM settings and plot the three AGN strengths (x10, x1, and x0) in order to span a full range of \lagn\ in our experiments.  Bins containing zero of these points are white, and bins with one or more point are shaded according to the median value of \lagn\ among the points it contains, as indicated by the colorbar: darker/blue regions have $\lagn \sim 1$, while lighter/yellow regions have $\lagn \lesssim 0.1$.  We plot histograms (on a logarithmic scale) along the edges of the first three rows to show the relative timescales of the model populations giving rise to these diagrams.  

We caution that our plotting and selection here are not meant to directly characterize the observed populations of ULIRGs that we plot in the right column; these are included primarily as a reference and plausibility check.   Instead, from this coding we seek to identify which regions, if any, isolate powerful AGN during these mergers, regardless of the length of time the observed properties may be visible.  A larger sample of models will be needed in order to fully constrain the expected observed distribution of \lagn\ at each location in these diagrams, but we can begin to highlight some basic trends implied by the simulations.  

\begin{deluxetable}{cccc}
\tablecaption{Selected Diagnostics, Deeply Obscured Case \label{tab:agnfrac}}
\tablehead{
\colhead{time range\tablenotemark{a}} & \colhead{$\log \pah$\tablenotemark{b}} & \colhead{$\log \tausi$\tablenotemark{b}} & \colhead{$\log \lagn$\tablenotemark{b}}
} 
\startdata
\sidehead{AGN x0, MW dust}
$t1 < t < t2$  & $  0.62$ & $ -0.31$ & $  0.00$ \\
$t2 < t < t3$  & $  0.56$ & $ -0.40$ & $  0.00$ \\
$t3 < t < t4$  & $  0.52$ & $ -0.37$ & $  0.00$ \\
\sidehead{AGN x1, MW dust}
$t1 < t < t2$  & $  0.57$ & $ -0.33$ & $  0.06$ \\
$t2 < t < t3$  & $  0.46$ & $ -0.62$ & $  0.33$ \\
$t3 < t < t4$  & $  0.39$ & $ -0.43$ & $  0.26$ \\
\sidehead{AGN x1, SMC dust}
$t1 < t < t2$  & $  0.12$ & $ -0.44$ & $  0.06$ \\
$t2 < t < t3$  & $ -0.13$ & $ -1.09$ & $  0.33$ \\
$t3 < t < t4$  & $ -0.03$ & $ -0.70$ & $  0.26$ \\
\sidehead{AGN x10, MW dust}
$t1 < t < t2$  & $  0.40$ & $ -0.37$ & $  0.43$ \\
$t2 < t < t3$  & $  0.04$ & $ -0.95$ & $  0.86$ \\
$t3 < t < t4$  & $  0.30$ & $ -0.40$ & $  0.81$ \\
\sidehead{Combined, MW dust}
$t1 < t < t2$  & $  0.57$ & $ -0.32$ & $  0.36$ \\
$t2 < t < t3$  & $  0.45$ & $ -0.65$ & $  0.70$ \\
$t3 < t < t4$  & $  0.39$ & $ -0.38$ & $  0.58$
\enddata
\tablenotetext{a}{These time ranges are 0.154, 0.056, and 0.252 $\times 10^9$ yr long, respectively }
\tablenotetext{b}{Median values}
\end{deluxetable}

\subsection{Comparison Between Models and Data} \label{ss:comparison}

\citet{veilleux09} analyzed the standard evolutionary scenario for ULIRGs \citep{sanders88a}, which broadly speaking proceeds through three stages.  In ``Stage 1'', the IR emission of a gas-rich merger is dominated by star formation, leading to large PAH EWs and relatively small silicate absorption EW.  As the merger proceeds to ``Stage 2'', $\tausi$ increases as the central source is buried in the gas-rich starburst, and the PAH emission decreases relative to the IR continuum that is boosted by the AGN, which remains sub-dominant to the starburst.  Finally, at ``Stage 3'', the silicate absorption and PAH emission both subside as the starburst fades, the obscuring medium is cleared or consumed, and the ULIRG is completely or nearly AGN dominated.  Overall, \citet{veilleux09} concluded that AGN contribute $\sim 35-40\%$ to the IR luminosity of ULIRGs, and that ULIRGs experience significant scatter from this general pattern.

In Figure~\ref{fig:data_comparison}, the highly obscured model points roughly fill the space spanned by observed sources.  Our procedure does not mandate such consistency, and so we interpret this as one important check of the modeling technique.  In particular, ignoring other diagnostics (such as SED slopes), this simulation reproduces the observed spread in $\tausi$ and $\pah$.  In Table~\ref{tab:agnfrac}, we summarize a few numerical results of our deeply obscured simulation.  We divide the merger into three stages, $t_1\thru t_2$ (pre-coalescence), $t_2\thru t_3$ (coalescence), and $t_3\thru t_4$ (post-coalescence), and report the median values of \lagn, \tausi, and \pah\ calculated in those periods.  

Generally, the SED properties along the model timeline follow those of the merger stages by V09, albeit with an expectedly high amount of scatter.  Local ULIRGs have a large amount of silicate absoprtion ($\tausi \sim 0.7$), suggesting they may correspond to the modeled points at $t_2 < t < t_3$ where $\tausi$ peaks.  However, the AGN fractions in models depend sensitively on the dust model and AGN strength choices, and so it is  unclear how well they may correspond to the $\sim 30\%$ reported in V09.  This value is consistent with our $\rm AGN\times 1$, MW dust simulation, but is much lower than the $\sim 70\%$ from the ``combined'' model set, in which each of the AGN strengths are given equal weight.  The more realistic choice is uncertain, but the $\rm AGN\times 1$ case has trouble reproducing observed objects with $\tausi < -1$.  However, SMC dust with lower AGN strengths can also provide that feature, and observations suggest that SMC-like dust may be more appropriate during bright starburst or AGN phases, as small grains are preferentially depleted \citep{Gordon:1997,Hopkins2004}.  This may imply that SMC dust is more appropriate to use in this situation; in any case, it highlights the difficulty in determining the path of the AGN fraction in such objects if their dust grain populations are changing significantly.

In Figure~\ref{fig:data_comparison}, swaths of strong and weak AGN exist, but most of these projections do not isolate AGN in simple ways.  For example, while it is true that sources with strong starbursts and weak AGN typically have stronger PAH emission and redder $\cone$ as discussed in Veilleux et al. (2009), strong AGN occupy all values of $\cone$, $\ctwo$, and $\tausi$.  Furthermore, the PAH strength depends sensitively on the dust model we assumed (see Figure~\ref{fig:time_evolution_dust}), and the $\pah$ values of low \lagn\ points depend on the obscuration level of the simulation.


In addition, we compare the relative fluxes emerging at hard X-ray and mid-IR wavelengths in the bottom row of Figure~\ref{fig:data_comparison}.  We compute model X-ray fluxes as in Section~\ref{ss:xraymethods}.  \citet{bauer10} found that $z > 1$ ULIRGs that are otherwise thought to be AGN-dominated (with low $\pah$) are undetected at X-ray wavelengths, suggesting that they are at least mildly Compton-thick.  Moreover, they presented Compton-thick examples from the literature at $z \sim 0$, which lie $\sim 2$ dex below the relation for less obscured AGN \citep[c.f.,][]{Honig2010}.  Many model galaxies with powerful AGN lie in the same region of this diagram, and are reflected in the dark blue squares mixed in with the lighter squares at $L_6 \sim 10^{11}\text{--}10^{12} L_{\odot}$.  These are the same sources in the low-$\pah$ (high $\tausi$) tail of dark points in the upper panels, implying that this class of Compton-thick sources is a short-lived phase of our highly obscured model.  The bulk of the simulation points have X-ray fluxes $\sim 0.5$ dex below the gray Type 1 AGN band, reflecting our choice to not analyze completely unobscured models.  

There are some properties for which the model values are not realized by observed sources.  Generally speaking, the differences are most apparent in the mid-IR SED slopes $\cone$ and $\ctwo$.  The most AGN-dominant model points are $\approx 0.5$ dex redder in $\ctwo$ than any observed object.  In bulk, the model $\cone$ values are larger by $\approx 0.3$ dex than the observed distribution.  And there are few model galaxies with $\log \tausi \sim 0$ \emph{and} $\log \pah \sim 0$, in contrast to those observed.  However, this is sensitive to our dust grain model (Section~\ref{ss:dustmodel}), and may also reflect our not having modeled AGN with no galaxy obscuration.  Our calculations are necessarily uncertain, so in Section~\ref{ss:improvements} below, we summarize some of these theoretical limitations and potential implications of these mismatches.  

As we have seen in Section~\ref{s:results}, when sufficiently obscured, a powerful AGN has an SED shape the same as a dusty pure starburst \citep[see also][]{Younger:2009,Narayanan:2010dog}.  Figure~\ref{fig:data_comparison} shows high \lagn\ squares occupying a large fraction of the $\pah$, $\tausi$, $\cone$, and $\ctwo$ values.  We also saw that the model SEDs can change drastically on timescales $\sim 10^{8}\rm\ yr$.  A further challenge is that the implied \lagn, given a set of SED properties, depends sensitively on assumptions about the dust model and AGN strength.  An ideal indicator is one that identifies powerful AGN regardless of the host galaxy's content or state.  An example is hard X-ray emission which can penetrate even very large optical depths, and in Section~\ref{s:implications}, we use our simulations to suggest other ideal indicators that have not been easily accessible with current data, but may be useful for future \mir\ studies with the \jwstfull\ (\jwst).

      \begin{figure}
	\begin{center}
	\includegraphics[width=3.2in]{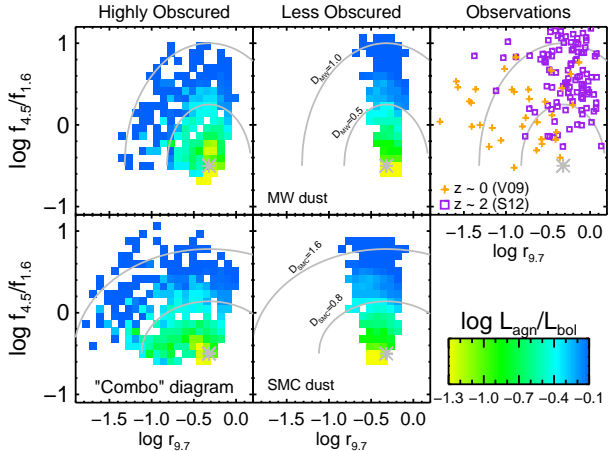}
	\end{center}
      \caption{ Following Figure~\ref{fig:data_comparison}, we plot panels motivating the diagnostic $D$: see text for definitions of $D_{\rm MW}$ and $D_{\rm SMC}$.  The left two columns are our simulated points, colored by \lagn, for our two representative levels of obscuration.  For this plot we include the AGNx10, AGNx1, and AGNx0 strengths.  The top row assumes MW dust, and the bottom assumes SMC dust.  In contrast to the other quantities from Figure \ref{fig:data_comparison}, diagnostic $D$ straightforwardly estimates \lagn\ for both low and high optical depth, assuming we know the dust model.  We note that diagnostic $D$ requires information at rest-frame wavelengths $1\thru10\mum$, rendering it somewhat impractical to measure using any single current instrument.  However, diagnostic $D$ will be accessible to the \jwstfull\ for sources at $z \lesssim 1$.   \label{fig:combo_diagram}}
      \end{figure} 


\section{Implications} \label{s:implications}

\subsection{An Ideal Indicator of \lagn} \label{ss:idealindicators}

Ultimately, we seek an indicator that can be used to determine the AGN contribution to IR-luminous galaxies across the full range of obscuration by galaxy dust.  Such a tool could simplify efforts to accurately track the growth of SMBHs and assess their impact on the ISM in diverse and evolving host galaxies.  

Near-infrared colors effectively identify AGN when their IR SED resembles a power-law, leading to enhanced emission beyond the stellar bump at $\gtrsim 2\mum$.  This yields a high value of our example near-IR color $\cfour$, for which the numerator is a proxy for torus emission and the denominator is a proxy for stellar mass.  Such techniques have been used to successfully identify high-redshift AGN-dominated ULIRGs using Wide-field Infrared Survey Explorer (WISE) photometry \citep[e.g.,][]{stern12,Eisenhardt2012,Wu2012}.  This near-IR reddening operates in powerful AGN for a wide variety of attenuation.  However, as we saw in Figure~\ref{fig:torus}, on short timescales the near-IR SED shape can change from a power-law AGN to one that resembles a starburst-dominated ULIRG when the obscuration increases drastically.  In these high-obscuration situations, $\tausi$ is sensitive to the AGN power, and so combining these indicators into a generic diagnostic may be feasible.  

In Figure~\ref{fig:combo_diagram}, we plot two-dimensional diagrams with our near-IR color $\cfour$ on the y-axis and $\tausi$ on the x-axis.  We have included our fiducial ISM assumptions for each mergers and all three AGN strengths: AGNx10, AGNx1, and AGNx0.  We note that these settings are chosen by hand in order to more broadly span the range of \lagn\ produced by our calculations, and are justified by the uncertainty inherent in the simulation methods and limited range of cosmologically motivated models we consider (see e.g., Section~\ref{ss:galaxymodels}).  Therefore, these model diagrams do not directly correspond with any realistic ULIRG populations; we plot two samples in the right panel primarily as a reference.

The light/yellow shaded region corresponds to low AGN emission ($\lagn \lesssim 0.1$) and is confined to a small region ($\sim 0.4$ by $0.4$ dex) of this parameter space centered at $(\log \tausi, \log \cfour) \approx (-0.32, -0.5)$.  In both the highly and less obscured simulations, the value $\log \cfour \sim -0.5$ corresponds to the SED shape of purely stellar emission, which can be seen in the gray dotted curves in Figure~\ref{fig:seds} at $1\mum < \lambda < 5\mum$.   The value $\log \tausi \sim -0.32$ appears to be the typical SED produced by ordinary (non-bursting) star formation in our simulations.  Numerically, this value may be set by our simplistic continuum estimate (Section~\ref{ss:simdiagnostics}), which is likely affected to some extent by the nearby PAH features.  Both of these ``zeropoints'' do not depend strongly on the dust model, and are roughly the same in the highly and less obscured mergers.  

Away from this point, the median AGN fraction varies in a semi-regular pattern.  Redder $\cfour$ implies a contribution at $4.5\mum$ from the relatively brighter AGN source, as expected from conventional hot dust/torus interpretations.  The absorption diagnostic $\tausi$ does not vary strongly in the less obscured case, as we have seen in Section~\ref{ss:agn}, and so the locus of increasing \lagn\ traces roughly a vertical line in the second column of Figure~\ref{fig:combo_diagram}.  

By contrast, in the highly obscured simulation, a significant area of the $\log \tausi \lesssim -0.5$ regions of the diagram is occupied by squares with $\lagn \gtrsim 0.1$.  In the lower left quadrant of each panel in the left column of Figure~\ref{fig:combo_diagram}, the dust column is large enough to produce a significant silicate absorption feature and to attenuate the near-IR continuum associated with AGN-heated dust, leading to $\log \cfour \lesssim 0$.  In the upper left quadrant, there are several squares with both a strong silicate feature and a red near-IR continuum.  At a fixed \lagn, these two quantities are correlated in the sense that increased silicate absorption implies more attenuation of the near-IR AGN-heated continuum relative to the starlight, giving a bluer $\cfour$, leading to the slanted or semi-elliptical pattern in the areas of constant shading (\lagn) in Figure~\ref{fig:combo_diagram}.  These qualities are the same when switching to the ``alt.\ ISM'' model for the highly obscured merger (not shown).  

We seek to summarize these trends and estimate \lagn\ directly from these variables.  Visually, the areas of constant \lagn\ form a semi-elliptical pattern that can be approximated by an analytic formula to potentially estimate \lagn\ with low scatter.  For MW dust, we consider the formula 
\[  D_{\rm MW} =  \sqrt{ \left (\log \tausi + 0.32 \right )^2 + \left ( \left [\log \frac{f_{4.5}}{f_{1.6}} + 0.5 \right ] \frac{2}{3} \right )^2}. \]
The effectiveness of combination $D_{MW}$ is evident in the top row of Figure~\ref{fig:combo_diagram}: the map between color/shading (\lagn) and these axes, $\cfour$ and $\tausi$, appears to be a well defined and slowly varying function of few parameters.  In Appendix~\ref{appendix}, we confirm the rough behavior of this diagnostic from toy models.  

We caution that we have analyzed only a handful of specially chosen models: AGNx10, AGNx1, and AGNx0 in two mergers that likely occupy two very different extremes of the ULIRG population.  Therefore the shaded points in Figures~\ref{fig:data_comparison} and \ref{fig:combo_diagram} are likely to change with a larger and more representative sample of models.  Nevertheless, it is encouraging that a clear trend in the $\cfour$--$\tausi$ plane exists for both our highly and less obscured mergers (in contrast to the diagrams in Figure~\ref{fig:data_comparison}), and for either the default or alternate ISM model in the highly obscured merger.  This can be understood as a result of the relatively simple ways that the two axes of Figure~\ref{fig:combo_diagram} reflect AGN-associated emission and its obscuration.  In Section~\ref{ss:dustmodel}, we show that this visual trend we identify in Figure~\ref{fig:combo_diagram} defines a tight relationship between $\lagn$ and $D_{\rm MW}$ at all times $t_1\thru t_4$ when the dust model is constrained.  

\subsection{Constraining Dust Properties} \label{ss:dustmodel}

However, we note that shape of the trend in the $\cfour$--$\tausi$ plane depends on our dust model: the bottom row of Figure~\ref{fig:combo_diagram} shows the same simulation using SMC dust instead of MW dust.  Importantly, a nice mapping in the above sense still exists, it is just skewed toward higher values of $\tausi$.  Therefore we can define a similar diagnostic,
\[  D_{\rm SMC} = \frac{5}{8} \sqrt{ \left (\log \tausi + 0.32 \right )^2 + \left ( \left [\log \frac{f_{4.5}}{f_{1.6}} + 0.5 \right ] \frac{5}{4} \right )^2}, \]
centered around the same point as $D_{\rm MW}$, but with a different axis ratio and scaling.

      \begin{figure}
	\begin{center}
	\includegraphics[width=3.2in]{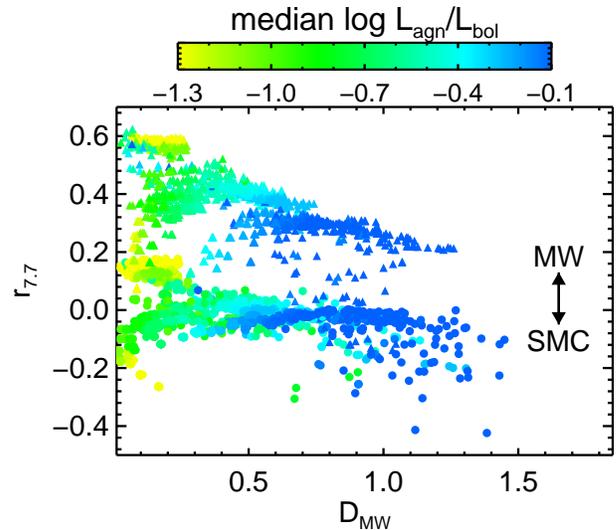}
	\end{center}
     \caption{ PAH strength versus our combination diagnostic $D_{\rm MW}$, which contains information about obscuration $\tausi$ and near-IR color, providing a rough estimate of \lagn\ at both low and high \mir\ optical depths.  If we have incorrectly assumed MW dust when the appropriate model is more like the SMC dust, then Figure~\ref{fig:combo_diagram} implies that the $D_{\rm MW}\thru\lagn$ relation will be skewed and have a large scatter.  However, at fixed high $D_{\rm MW}$, the PAH strength can roughly select the appropriate dust model, and therefore the appropriate predictor of \lagn: $D_{\rm MW}$, $D_{\rm SMC}$, or an interpolation.   For this plot we have included the AGNx1 and AGNx10 strengths for the default ISM model of the highly obscured merger.  \label{fig:ir_dust} }
      \end{figure} 

In order to select between these estimators, we must first constrain the dust properties; in other words, we must jointly constrain the dust and \lagn.  One option is to use the PAH emission: assuming SMC dust, $\pah \sim 0$ for all model points instead of $\sim 0.3$ for MW dust (see Figure~\ref{fig:time_evolution_dust}).  We demonstrate this procedure in Figure~\ref{fig:ir_dust}: while $D_{\rm MW}$ does not yield a low-scatter predictor of \lagn\ when applied to a system with SMC dust, it does broadly select stronger AGN.  Moreover, at $D \gtrsim 0.2$, sources with these different dust models separate based on their PAH strength, $\pah$, permitting either a direct choice between the formulae for $D_{\rm MW}$ and $D_{\rm SMC}$ or an intermediate case.  This suggests the following procedure:
\begin{enumerate}
\item{Estimate $D_{\rm MW}$.}
\item{Estimate the PAH strength (here, $\pah$).}
\item{Select dust model X based on the value obtained in step 2, by e.g., checking whether at the $D_{\rm MW}$ value, $\pah$ lies closer to the MW or SMC points.}
\item{Calculate $D_{\rm X}$ and use it to estimate \lagn.}
\end{enumerate}

\begin{figure*}
	\begin{center}
	\begin{tabular}{c}
	\includegraphics[width=7.0in]{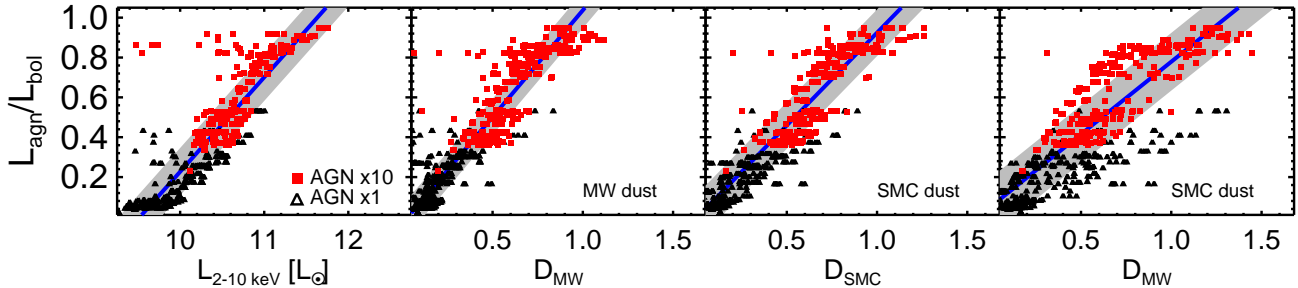}
	\end{tabular}
	\end{center}
      \caption{ We show how diagnostic $D$ correlates with \lagn\ in our simulations in the right three panels.  We show here only the highly obscured merger with our default ISM model, but the trends are similar for the alternate ISM model and the less obscured merger, as summarized in Table~\ref{tab:lagnpredict}.  As a comparison, we plot the hard X-ray flux through the $2-10\rm\ keV$ band in the left panel.   If the correct dust model is used, D accurately predicts \lagn\, as well as \lhard.  In the rightmost panel, we apply $D_{\rm MW}$ when the source has SMC dust instead, demonstrating that the scatter increases by $\sim 70\%$.  This issue can be mitigated by using, for example, Figure~\ref{fig:ir_dust} to choose the appropriate indicator D.  \label{fig:D_v_lagn} }
      \end{figure*}

With a known dust model, diagnostic $D$ robustly predicts $\lagn$ regardless of the physical conditions of the ULIRG source.  We compare the ability of diagnostics $D_{\rm MW}$ and $D_{\rm SMC}$ to predict $\lagn$ to that of hard X-ray fluxes in Figure~\ref{fig:D_v_lagn}.

Using a least absolute deviation method, we fit the relation $\lagn = a + b Q$ for three quantities $Q$ ($D_{\rm MW}$ with MW dust, and both $D_{\rm MW}$ and $D_{\rm SMC}$ with SMC dust) and each of our two ISM models for the highly obscured simulation, and compile the results in Table~\ref{tab:lagnpredict}.  To measure the scatter in the correlations' residuals, we used the same scale estimator $S$ defined in Section \ref{ss:angle}, which is generally insensitive to outliers.  In both cases, $\lhard$ and each $D$ predict $\lagn$ with roughly the same median scatter, $S \sim 10\%$.  However, we find that $\lhard$ experiences a failure rate ($> 3\sigma$) two to three times higher than $D$ in the highly obscured calculation.  Both quantities predict $\lagn$ equally well under the alternate ISM treatment and for the less obscured candidate (not shown), with $S < 10\%$ and failure rates $< 1\%$.  

Several simulated points at $\lagn \approx 0.8$ fail in both \lhard\ and $D_{\rm MW}$.  In these cases, corresponding to the most highly obscured lines of sight in our $\rm AGN\times10$ calculation from Figure~\ref{fig:cameraseds}, the \mir\ is so heavily attenuated that no \tausi\ signature from the central source survives in the observed SED (see also Section~\ref{ss:angle}).  

\begin{figure}
	\begin{center}
	\includegraphics[width=3.2in]{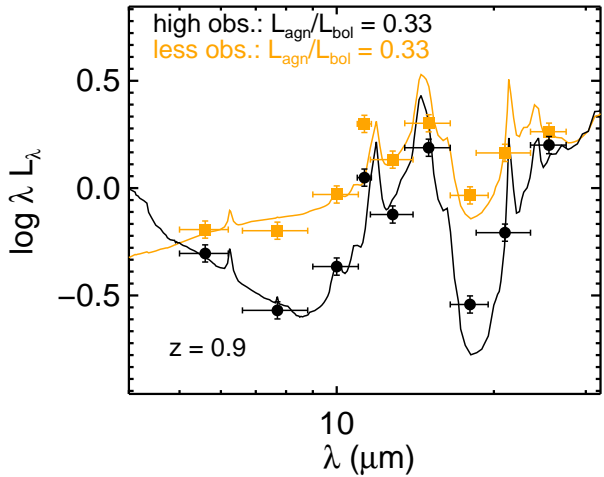}
	\end{center}
      \caption{ Synthetic photometric observations by MIRI on \jwst, demonstrating the ability of future \mir\ technology to measure AGN power in systems with a wide range of obscuration.   Here we assumed MW dust and $z=0.9$.  \label{fig:datased}}
      \end{figure}

Diagnostic $D$ requires information about the galaxy's stellar mass (rest-frame $\lambda \sim 1\thru2\mum$), limiting its immediate application to large new samples.  However, the rest-frame wavelength range $D$ requires will be covered by the \jwstfull\ (\jwst), an observatory whose Mid-Infrared Instrument (MIRI) \citep[e.g.,][]{wright04_miri} permits measurements of $\pah$ and $\tausi$ for sources at $z \lesssim 1$.  The tight correlation we find between $D$ and $\lagn$ reflects the possible benefit in combining information about $\tausi$ and $\pah$ with near-IR color diagnotics from \jwst\ itself \citep[e.g.,][]{messias12}, or by matching to sources from other IR surveys \citep[e.g.,][]{stern05,stern12}.  For the two example SEDs in Figure~\ref{fig:datased}, whose known AGN fractions are both $\lagn = 0.33$, this procedure estimates $\lagn = 0.33 \pm 0.09$ and $\lagn = 0.38 \pm 0.09$ (uncertainties are from the minimum theoretical scatter in the correlations) for the high and low obscuration examples, respectively.

\subsection{Limitations and Improvements} \label{ss:improvements} \label{ss:limitations}

A concern with the present method is that we cannot be sure how our few idealized simulations fit into a cosmological context.  We have mitigated this issue by presenting two rather different ULIRG candidates, by varying the relative power between the AGN and star formation, and by varying the assumptions acting on dust obscuration and emission.  We are planning to build a library of model SEDs for a much wider variety of systems, enabling more robust studies across IR-luminous galaxy populations.  Moreover, we have focused here on AGN signatures at mid-infrared wavelengths, and information at longer wavelengths \citep[c.f.,][]{Younger:2009}, such as the location of the far-IR peak, could further constrain the AGN fractions in such sources (e.g., S12).  In an upcoming paper (E.~Roebuck et al.\ in prep.), we include the far-IR peak predictions from these simulations in interpreting the properties of $z\sim0.3-3$ starbursts and quasars (the S12 sample).

The simplest interpretation of Figure~\ref{fig:data_comparison} is that analogues of our ``highly obscured'' and ``less obscured'' examples are both realized in observed sources, in addition to both MW and SMC dust: all four models are required to cover the wide observed distributions of \mir\ color, PAH emission, and silicate absorption.  A higher ratio of silicate to carbonaceous dust grains increases $\tausi$ while shrinking $\pah$ (Figures~\ref{fig:time_evolution_dust}), without changing $\cfour$ (Figure~\ref{fig:combo_diagram}), permitting estimates of \lagn\ under a variety of physical conditions, dust models, and optical depths.   Although different ISM or dust models can produce broadly similar \mir\ trends, a separate measure of the AGN power, such as X-ray flux, is essential to further constrain these models.  Furthermore, the recent discovery of high-redshift, hyper-luminous, hot-dust ULIRGs \citep{Eisenhardt2012,Wu2012} suggests that we may lack a key phase in these few simulations, in which feedback initiates a transition between scenarios represented by our ``highly obscured'' and ``less obscured'' SED models.  

Several assumptions regarding the dust radiative transfer may lead to discrepancies between our model sources and real sources.  In particular, from Figure~\ref{fig:data_comparison} we have identified two failures to which these assumptions likely contribute: 1) the \mir\ colors of the most AGN-dominated highly obscured sources are redder than their real counterparts (although bluer at $\lambda \lesssim 5 \mum$; see Figure~\ref{fig:combo_diagram}), and 2) we do not simulate enough sources with little signs of PAH emission or silicate absorption: $\tausi \sim 0$ \emph{and} $\pah \sim 0$.

The IR calculation in \sunrise\ distributes the energy absorbed by small PAH grains equally between emission in thermal equilibrium and in modes giving rise to the distinctive \mir\ PAH lines \citep{Jonsson:2010sunrise}.  Therefore, any change to this assumption or more sophisticated calculation, such as thermally fluctuating small grains \citep[e.g.,][]{Guhathakurta1989}, will affect the dust continuum shape and PAH strength.  In addition, we do not currently model dust destruction, and our model for dust production is crude, both key areas where near-future progress may be tenable.  By not allowing our radiation to destroy dust grains or alter the dust grain distribution (or changing more generally the gas distribution owing to realistic feedback), we may be preventing ourselves from obtaining truly ``power-law'' AGN SEDs with no \mir\ dust features that may arise after AGN feedback disrupts the ISM.

We have purposefully focused on how galaxy dust can affect the fixed SMBH source SED (with minor variations in Section~\ref{ss:torus}), rather than how the physics on smaller scales play out in observations of ULIRGs.  While high levels of galaxy dust can reproduce a number of properties of obscured quasars, the evolution of the central regions during these times of maximum obscuration will depend on the behavior of gas structures very close to the SMBH.  This behavior may set the minimum scatter to be observed in correlations between these indicators.  Using multiscale simulations, \citet{Hopkins2012a} showed how gravitational instabilities propagate down to scales of $\sim 1\thru 100$ pc during large-scale gas inflows.  They demonstrated that a number of properties of obscured AGN attributed to a torus are explained by such dynamically-evolving accretion structures.  Furthermore, these simulations suggest that the angular momentum vector of the small-scale ``torus'' can often be misaligned with the angular momentum of gas on galaxy scales, and the relative orientations can vary significantly in time \citep{Hopkins2012b}. Thus, the attenuation from the host galaxy may vary more significantly than our current simulations suggest.  Therefore we are motivated to develop methods of integrating knowledge of the impact of these different scales on the emergent SEDs as a means for further interpreting the evolution of IR-luminous galaxies.

Recent work on merger simulations at higher resolution \citep{Hopkins2013} and with different hydrodynamics solvers (e.g., \arepo: \citealt{Springel2010}, C.\ Hayward et al.\ in prep.) shows that the average global evolution in simulations like the ones we analyze here, and hence conclusions made based on them, are robust.  However, such advances will enable more sophisticated treatments of feedback processes and hydrodynamics in gas with a complex phase structure and at smaller scales; therefore future SED modeling could exploit these improvements to produce ever more detailed comparison libraries.  



\begin{deluxetable}{ccccc}
\tablecaption{Ability to Predict $\lagn$ \label{tab:lagnpredict}}
\tablehead{
\colhead{Quantity, $Q$\tablenotemark{a}} & \colhead{scatter, $S$\tablenotemark{b}} & \colhead{intercept, $a$} & \colhead{slope, $b$} & \colhead{Failure $\%$\tablenotemark{c}}
} 
\startdata
\sidehead{Default ISM}
$\log L_{2-10\rm\ keV}/L_{\odot}$ & $0.12$ & $-4.51$ & $0.47$ & $6.1$ \\
$D_{\rm MW}$, MW dust & $0.09$ & $-0.04$ & $1.07$ & $2.6$ \\
$D_{\rm SMC}$, SMC dust & $0.12$ & $-0.01$ & $0.93$ & $0.5$ \\
$D_{\rm MW}$, SMC dust & $0.14$ & $0.05$ & $0.72$ & $0.8$ \\

\sidehead{Alternate ISM}
$\log L_{2-10\rm\ keV}/L_{\odot}$ & $0.09$ & $-4.59$ & $0.47$  & $0.8$ \\
$D_{\rm MW}$, MW dust & $0.10$ & $-0.04$ & $0.96$ & $0.9$ \\
$D_{\rm SMC}$, SMC dust & $0.09$ & $0.00$ & $0.86$ & $0.0$ \\
$D_{\rm MW}$, SMC dust & $0.10$ & $0.00$ & $0.98$ & $1.3$ \\
\enddata
\tablenotetext{a}{$\lagn = a + b\left ( Q\right )$}
\tablenotetext{b}{$S = $ MAD$/0.67$, see Section~\ref{ss:angle}.}
\tablenotetext{c}{Percentage of points with $\lagn$ further than 3$\sigma$ away from the fitted relation.  In practice, for these cases the indicator predicts an $\lagn$ smaller than the true value owing to extreme attenuation.}
\end{deluxetable}

\section{Summary and Conclusions} \label{s:conclusions}

In this work we used the detailed information contained in high-resolution hydrodynamical merger simulations to analyze the impact of galaxy-scale evolution of dust on interpretations of AGN power in ULIRGs derived from the mid-infrared.  Using the dust radiative transfer code \sunrise, we calculated the SED evolution during two model starbursts meant to represent possible extremes of observed ULIRG populations.  One merger is characteristic of bright starbursts at $z\sim 2\thru 4$, while the other is a typical gas-rich merger at $z \sim 0$. 

Increasing computational resources will enable future analyses of many more galaxy and galaxy merger scenarios and more sophisticated modeling of the small-scale physics that are presently very uncertain.  In this initial work, we studied the variation of the modeled \mir\ SEDs as a function of viewing angle, source SED, dust grain distribution, galaxy ISM clumpiness, and AGN power, and evaluated the ability of common mid-IR diagnostics to predict the AGN UV-IR luminosity fraction, \lagn.  We developed a possibly generic \mir\ estimator of \lagn\ that can be measured for new samples with next-generation instruments such as \jwst.

Our main conclusions are:
\begin{enumerate}
\item{Our modeling technique broadly reproduces the observed span of common \mir\ diagnostics used to disentangle AGN and starburst activity.}
\item{Although generally consistent with previous interpretations, none of the indicators straightforwardly predict \lagn\ because they depend non-linearly on the galaxy's properties, which can vary rapidly on timescales $\sim 10^{8}\rm\ yr$.  }
\item{Highly obscured sources can be more AGN-dominated than realized, even when the shape of their SED resembles a pure starburst.  A large enough column density reprocesses all of the direct AGN light into longer wavelengths.  }
\item{The mid-IR SEDs of sources with a powerful, highly obscured AGN depend strongly on the direction from which the source is observed.  Along some lines of sight, significant dust self-absorption attenuates the IR flux and obscures the mid-IR signatures of the buried AGN.}
\item{Sources with a deep $9.7\mum$ silicate absorption feature can be reproduced by models either with an extremely luminous AGN and MW galaxy dust, or with an average-strength AGN and SMC galaxy dust. }
\item{It is possible to construct an idealized \mir\ indicator utilizing future \jwst\ data to cleanly estimate \lagn\ while simultaneously constraining the dust grain model.}
\end{enumerate}

\acknowledgements

This research has made use of NASA's Astrophysics Data System.  The computations in this paper were run on the Odyssey cluster supported by the FAS Science Division Research Computing Group at Harvard University.  We thank the anonymous referee for valuable suggestions that improved this paper.  


\appendix

\section{Toy Models for Qualitative Diagnostic Behavior}  \label{appendix}

Here we make plausibility arguments based on a simple toy model to validate the simulated behavior of the diagnostics we analyzed. We assume a point source AGN that has an SED composed of direct AGN emission and dust re-emission from the torus.  We label this flux $c_{\rm A}$.  This is attenuated by a foreground screen such that the flux at infinity is $c_{\rm A} e^{- \tau}$.  For the starburst component, $c_{\rm S}$, energy is absorbed by dust and re-emitted at IR wavelengths.  Then, this dust emission is attenuated at larger scales with some effective optical depth.

The presence of silicate grains causes a sharp increase in the dust opacity near $ \lambda = 9.7 \mum$.   Suppose that the attenuation curve is such that the attenuation on either side of the deep silicate feature is $e^{-\tau_{9.7}}$, but in the feature, the attenuation is $e^{-R\tau_{9.7}}$ where $R > 1$.  $R$ represents the ratio of the peak of the full extinction curve at $9.7 \mum$ to that arising without the silicate absorption; here we analyze $R = 6$, which roughly matches the behavior we see in the simulations as expected with an extinction curve for MW-like dust \citep{Li2001}.  


As in Section~\ref{ss:simdiagnostics}, $\tausi = f_{9.7}/c_{9.7}$. We define the intrinsic monochromatic AGN fraction $F_{\rm A} = {c_{\rm A}}/({c_{\rm A} + c_{\rm S}})$.  This quantity is closely related to, but not directly proportional to, the bolometric AGN fraction \lagn\ used throughout the rest of this paper and defined in Section~\ref{ss:agnfractiondefinition}.  We estimate the conversion between these quantities in Equation~\ref{eq:bolcorrect}.  By deriving toy diagnostics as a function of $F_A$ we can controllably analyze their behavior in much the same way that we explored the simulations in Sections~\ref{s:results}, \ref{s:comparison}, and \ref{s:implications} using the AGNx0, AGNx1, and AGNx10 models defined in Sections~\ref{ss:alternatemodels} and \ref{ss:galaxymodels}.  Expanding \tausi, we find,
\begin{equation}  \label{eq:tausi}
\tausi = \frac{f_{9.7}}{c_{9.7}} = \frac{c_{\rm A} e^{-R\tau_{9.7}} + a c_{\rm S} }{c_{\rm A} e^{-\tau_{9.7}} + c_{\rm S}} = \frac{F_{\rm A} e^{-R\tau_{9.7}} + a (1-F_{A})}{F_{\rm A} e^{-\tau_{9.7}} + (1-F_{A})} , 
\end{equation}
where $a$ is a new parameter that characterizes the silicate absorption feature imprinted on the IR energy emitted from star formation.  In the RT calculations, we obtain $a \sim 0.5$ since $\log_{10} \tausi \sim -0.3$ when $F_{\rm A} \sim 0$ (Figure~\ref{fig:time_evolution});  this corresponds to an effective optical depth to the star forming regions of $\tau_{9.7,\rm SF} \sim 0.13$.  Since the SF gas is spatially extended, $\tau_{\rm SF}$ will change less with viewing angle than $\tau_{\rm A}$, allowing us to keep this parameter $a$ fixed.  

The PAH fraction $\pah$ is defined to be $\pah = f_{7.7}/c_{7.7}$.  Add new variables $p_{\rm A}$ and $p_{\rm S}$ for the PAH emission luminosity, and assume that the PAH emission is some fixed factor $P$ times the continuum emission for either AGN or star formation.  We expand to find,
\begin{equation} \label{eq:pah}
\pah = \frac{(c_{\rm A} + p_{\rm A})e^{-\tau_{7.7}} + (c_{\rm S} + p_{\rm S})}{c_{\rm A}e^{-\tau_{7.7}} + c_{\rm S}} \approx \frac{(1 + P)c_{\rm A}e^{-\tau_{7.7}} + (1 + P)c_{\rm S}}{c_{\rm A}e^{-\tau_{7.7}} + c_{\rm S}} \sim 1+P .
\end{equation}
The true behavior is somewhat more complex than this, since in Figure~\ref{fig:time_evolution} we see that $\pah$ varies quickly at high AGN fractions and obscuration levels, but since both the numerator and denominator have a similar dependence on $\tau_{7.7}$, there is minimal scatter with viewing angle at all times.  

Now, if we ignore SF by setting $F_{\rm A} = 1$, we get $\tausi = e^{-(R-1)\tau_{9.7}}$, or $\ln \tausi = -(R-1)\tau_{9.7}$ for all $\tau_{9.7}$.    The magnitude of this pre-factor ($R-1 > 1$) implies that the depth of the silicate feature varies more strongly with optical depth, and thus viewing angle and time, than does the continuum.  Its negative sign gives rise to the intuitive behavior that the silicate absorption feature becomes more apparent ($\tausi$ shrinks) with increasing optical depth.  However, this trend fails in, e.g., Figure 6, and the D points scattered low in Figure~\ref{fig:D_v_lagn}.  

For simplicity, in the case of mixed SF and AGN, let $a = 0.5$ and $F_{\rm A} = 0.9$, but let $\tau$ vary.  When $\tau$ is very small, then the AGN terms dominate, yielding $ \ln \tausi \sim -(R-1)\tau, \ \ \tau \ll 1$, the same behavior as the no-SF case.  When $\tau$ is large enough that the AGN term in the numerator of Equation~\ref{eq:tausi} can be safely ignored, then we have $\tausi \approx {0.5}/{(9 e^{-\tau} + 1)} $. This simplifies depending on $\tau$:

$$\ln \tausi \approx  - \ln 9 +  \tau,\ \ \tau \sim 1, \quad \quad \ln \tausi \approx - 9 e^{-\tau},\ \ \tau \gg 1$$

\begin{figure}
	\begin{center}
	\includegraphics[width=6.5in]{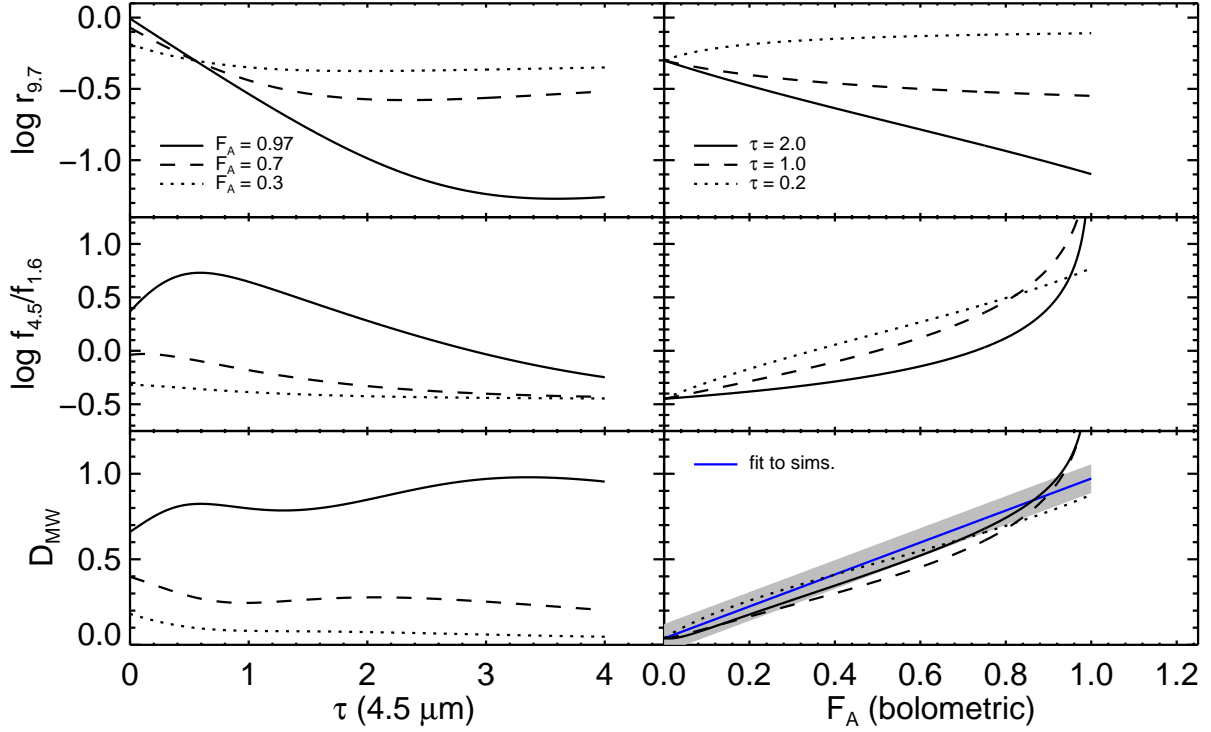}
	\end{center}
      \caption{ Behavior of toy models for $\tausi$, $\cfour$, and $D_{\rm MW}$ as a function of near-IR optical depth and AGN fraction.  These simple models consist of an AGN source buried behind a simple obscuring screen (``galaxy dust''), plus an extended region of star formation that has a fixed spectral shape that does not depend on the optical depth to the AGN.  In the upper left panel, $\log \tausi$ has a strong dependence on $\tau$, even at low levels of obscuration, giving rise to the large viewing angle scatter of $\tausi$ in the simulations of Section~\ref{ss:angle}: since the AGN is a point source, its final SED depends only on attenuation along a single line of sight, which can vary widely, while indicators of activity over a larger spatial region (such as $\pah$) have their attenuation averaged over many lines of sight.  Several other phenomena from the simulations are demonstrated by these models:  1) Generally, the silicate feature is enhanced as the AGN fraction increases and as $\tau$ increases, up until the SF gas begins to dominate the attenuated AGN SED, at which point the dependence of $\tausi$ on $\tau$ reverses; 2) The near-IR color $\cfour$ effectively selects powerful AGN when $\tau$ is small enough (in our units, $\tau (4.5\mum) \lesssim 1\thru 2$); and 3) At larger $\tau$, $\log \cfour$ decreases and $| \log \tausi |$ increases in such a way that the combination of these diagnostics, $D_{\rm MW}$, remains roughly constant for a fixed AGN fraction.  This constancy is shown in the bottom left panel, and gives rise to the tight correlation between AGN fraction and $D_{\rm MW}$ in the bottom right panel.  These simple arguments roughly verify the RT calculations and motivate the behavior of the empirically defined indicator $D_{\rm MW}$ (Section~\ref{ss:idealindicators}).     \label{fig:appendixfig}}
      \end{figure}

Therefore, for powerful AGN, $\tausi$ depends strongly on $\tau$, until $\tau$ gets too far above 1, giving rise to the strong viewing-angle scatter we noticed in the simulations in Section~\ref{ss:angle}.  In Figure~\ref{fig:appendixfig}, we plot example curves showing how $\tausi$ and $\pah$ depend on $F_{\rm A}$ and $\tau$ in this toy model, which generally match the behavior we find in the simulations.  

The qualitative behavior of diagnostic $D_{\rm MW}$ can be derived in the same fashion.  For this, we seek a similar type of model for $\cfour$.   Roughly speaking (Figure~\ref{fig:diagram}), the pure-AGN SED is such that  $f_{4.5} = f_{1.6} \left ( \nu_{1.6}/\nu_{4.5} \right ) = 2.8 f_{1.6}$, and the stellar SED is $f_{4.5} = f_{1.6} \left ( \nu_{4.5}/\nu_{1.6} \right ) = f_{1.6}/2.8$.  The added complication in this situation is the attenuation of the AGN source depends on wavelength, roughly speaking as $\tau (\lambda) \sim \tau (1\mum)\times  (\lambda / 1\mum)^{1.5} $ (e.g., \citealt{Weingartner:2001}).  We can use the same correction when combining $\tausi$ and $\cfour$ to model the $D_{\rm MW}$ diagnostic.  We get: 
\begin{equation}
 \frac{f_{4.5}}{f_{1.6}} \approx \frac{F_{\rm A}e^{-\tau_{4.5}} + (1 - F_{\rm A})}{(F_{\rm A}/2.8)e^{-\tau_{4.5} 2.8^{1.5}} + 2.8(1 - F_{\rm A})}
\end{equation}
In Figure~\ref{fig:appendixfig}, we construct \cfour\ and $D_{\rm MW}$ analytically and plot their behavior as a function of $F_{\rm A}$ and $\tau$ (at 4.5\mum).  A final difference between the simulated quantities and these toy models is that we derived these models as a function of a monochromatic AGN fraction, while the quantity \lagn\ used in the rest of this paper is the bolometric AGN fraction.  The correction depends on the SED shape of the various components, which we parametrize as $P_{\rm A} = j P_{\rm A, bol}$, $P_{\rm SF} = k P_{\rm SF, bol}$.  The AGN bolometric correction factor $j$ is fixed since our input template has a fixed SED shape over the wavelengths used by \sunrise, while the SF correction $k$ varies; we find that $j/k \approx 5\pm 2$ in our RT calculations, and so we derive:
\begin{equation} \label{eq:bolcorrect}
F_{\rm A} = \frac{P_{\rm A}}{P_{\rm A} + P_{\rm SF}} = \frac{k P_{\rm A,bol}}{k P_{\rm A,bol} + j P_{\rm SF, bol}} = \frac{(k/j) F_{\rm A,bol}}{F_{\rm A,bol}(\frac{k}{j} - 1) + 1} \sim \frac{5 F_{\rm A,bol}}{4F_{\rm A,bol} + 1}.
\end{equation}  
Figure~\ref{fig:appendixfig} shows \tausi, \cfour, and $D_{\rm MW}$ from these idealized formulae for a sampling of AGN fractions and optical depths, confirming much of the behavior we observed in the results of the RT calculations.  

\end{document}